\documentclass[12pt]{article}

\usepackage{scicite}
\usepackage{graphics}

\usepackage{times}
\newenvironment{sciabstract}{%
\begin{quote} \bf}
{\end{quote}}

\newcounter{lastnote}

\title{The Shape of Saturn's Huygens Ringlet Viewed by Cassini ISS}

\author
{Joseph N. Spitale ,$^{1\ast}$ Joseph M. Hahn,$^{2}$\\
\\
\normalsize{$^{1}$Planetary Science Institute,}\\
\normalsize{1700 E. Ft. Lowell Ste 106, Tuson, AZ 85719, USA}\\
\\
\normalsize{$^{2}$Space Science Institute,c/o Center for Space Research, University of Texas at Austin,}\\
\normalsize{3925 West Braker Lane, Ste 200, Austin, TX 78759-5321, USA}\\
\\
\normalsize{$^\ast$To whom correspondence should be addressed; E-mail:  jnspitale@psi.edu.}
}

\date{}

\begin{document}

\baselineskip24pt

\maketitle

%=====================================================================
% {Abstract} 
%=====================================================================
\begin{sciabstract}
A new model for the shape of the prominent eccentric ringlet in the gap exterior to Saturn's B-ring is developed based on Cassini imaging observations taken over about 8 years.  Unlike previous treatments, the new model treats each edge of the ringlet separately.  The Keplerian component of the model is consistent with results derived from Voyager observations, and $m=2$ modes forced by the nearby Mimas 2:1 Lindblad resonance are seen.  Additionally, a free $m=2$ mode is seen on the outer edge of the ringlet.  Significant irregular structure that cannot be described using normal-mode analysis is seen on the ringlet edges as well.  Particularly on the inner edge, that structure remains coherent over multi-year intervals, moving at the local Keplerian rate.  We interpret the irregular structure as the signature of embedded massive bodies.  The long coherence time suggests the responsible bodies are concentrated near the edge of the ringlet.  Long wake-like structures originate from two locations on the inner edge of the ringlet, revealing the locations of the two most massive embedded bodies in that region.  As with the Voyager observations, the Cassini data sets showed no correlation between the width and the radius of the ringlet as would be expected for a self-gravitating configuration, except for a brief interval during late 2006, when the width-radius relation was similar to those seen in most other narrow eccentric ringlets in the Solar System.
\end{sciabstract}

%==========================================================
\section{Introduction} 
%==========================================================

%%--------------------------------------------------------------------
%\begin{figure}[h!]
% \begin{center}
%{\scalebox{0.25}{\includegraphics{./ringlets.eps}}}
% \caption{\small Comparison of Huygens ringlet with other Saturnian ringlets.  %Each frame encompasses about 1000 km of ring radius.}
% \label{fig:ringlets}
%\end{center}
%\end{figure}
%%--------------------------------------------------------------------

Saturn's Huygens ringlet, located $\sim$ 250 km exterior to the outer edge of Saturn's B ring, has for some time been known to be eccentric with a radial amplitude of $\sim$ 32 km \cite{Porco1983, Porco1990, Turtle1991}.  In Cassini images, the Huygens ringlet appears similar to the isolated C-ring ringlets Maxwell and Colombo, with well-defined edges and a normal optical depth around 0.5.  However, as will be shown in this paper, the overall shape of the Huygens ringlet bears more resemblance to the F ring, due to irregular radial deviations from its primary Keplerian shape.  

Based on the relatively large residuals in the simple Keplerian ellipse model, \cite{Porco1983}\footnote{In that work, the names of the Huygens and Maxwell ringlets were swapped.} hypothesized that an additional wave-number-2 (i.e., $m=2$) pattern may be present as a response to the strong Mimas 2:1 inner Lindblad resonance and/or the 74-km-amplitude\footnote{As measured in that work.} $m=2$ distortion in the massive B-ring edge, which was assumed to be produced by the resonance.  An attempt to fit such a pattern using 13 Voyager measurements (11 ISS images, one RSS occultation, and one UVS occultation) did not produce a statistically significant result.  However, shortly after the start of Cassini's prime mission, a wave-number-2 mode was detected with a radial amplitude of $\sim$ 2 km \cite{Spitale2006a}, too small to have been detected in the Voyager data sets.

Although the $m=2$ disturbance at the outer edge of the B ring was originally viewed as a static response to the Mimas resonance, recent work \cite{Hedman2010, Spitale2010} showed that the pattern is time-variable, and there is now strong evidence that the time variability arises from interference with an additional free $m=2$ mode whose amplitude is slightly larger than, and whose pattern speed is slightly faster than that of the forced mode \cite{Spitale2010}.  If the forced $m=2$ pattern in the B-ring could affect the Huygens ringlet, then this free $m=2$ mode might also produce an effect.  Moreover, the Huygens $m=2$ pattern might itself be such a free mode.  Therefore there are a number of plausible wave-number-2 patterns that might exist in the Huygens ringlet, and they can be distinguished by their pattern speeds.

In this work, we use methods similar to those employed in our investigations of the Saturn's A- and B-ring edges \cite{Spitale2009, Spitale2010} to model the kinematics of inner and outer edges of the Huygens ringlet using Cassini imaging data sets.  We develop refined parameters for the ringlet's $m=1$ Keplerian shape, and we search for modes with higher wave numbers.  We examine the ringlet's irregular structure, and we look at the ringlet's unusual width-radius relation.

%==========================================================
\section{Approach} 
%==========================================================

%--------------------------------------------------------------------
% data sets
%--------------------------------------------------------------------
\begin{table}[h!]
 \tiny
 \center
 \begin{tabular}{lllllll}
Set & Date & $\Theta_0$ ($^\circ$) & $\Theta$ ($^\circ$) & $g$ ($^\circ$) & Radial Scale & Observation ID \\ 
 &  &  &  &  & (km pixel$^{-1}$) & \\ \hline

2$^a$  & 2005-174 & 111 & 71--72 & 45--48   & 9.8--10.7  & ISS\_010RI\_LPMRLFMOV001\_PRIME \\
3$^b$  & 2005-176 & 111 & 65--70 & 20--29   & 5.7--21.3  & ISS\_010RI\_AZSCNLOPH001\_PRIME \\
5$^b$  & 2005-231 & 111 & 64--73 & 3--13    & 3.7--21.7  & ISS\_013RI\_AZSCNLOPH001\_PRIME \\
6$^a$  & 2006-247 & 106 & 85--85 & 163--164 & 12.5--13.1  & ISS\_028RI\_HIPHAMOVE001\_PRIME \\
8$^b$  & 2006-312 & 105 & 24--45 & 34--103  & 2.1--3.0  & ISS\_031RI\_WN60209001\_PRIME \\
9$^a$  & 2006-322 & 105 & 84--88 & 143--148 & 6.5--7.2  & ISS\_033RI\_BRINGBLUR003\_PRIME \\
20$^a$ & 2007-083 & 77  & 69--71 & 135--138 & 3.7--3.7  & ISS\_041RI\_BRINGBLUR015\_PRIME \\
23$^a$ & 2007-099 & 103 & 54--68 & 18--23   & 3.9--4.9  & ISS\_042RI\_LPHRLFMOV001\_PRIME \\
26$^a$ & 2007-118 & 103 & 80--81 & 46--47   & 9.7--10.0  & ISS\_043RI\_LPMRDFMOV001\_PRIME \\
43$^b$ & 2008-028 & 99  & 30--68 & 25--84   & 1.8--4.6  & ISS\_057RI\_AZSCAN001\_PRIME \\
44$^a$ & 2008-069 & 98  & 52--52 & 47--48   & 8.3--8.4  & ISS\_061RI\_BRINGBLUR001\_PRIME\\
46$^b$ & 2008-097 & 98  & 65--67 & 31--32   & 8.8--21.2  & ISS\_064OT\_RETARGEMR002\_PRIME \\
55$^a$ & 2008-112 & 83  & 40--45 & 64--72   & 2.8--3.1  & ISS\_065OT\_RETARGEMR004\_PRIME \\
31$^a$ & 2008-121 & 83  & 55--81 & 130--139 & 1.4--1.4  & ISS\_066RI\_RETARGHPH001\_PRIME\\
33$^b$ & 2008-151 & 97  & 63--69 & 31--35   & 6.9--30.5  & ISS\_070OT\_RETMDRESA020\_PRIME \\
34$^b$ & 2008-159 & 97  & 52--56 & 41--44   & 5.6--14.6  & ISS\_071OT\_PAZSCN002\_PRIME \\
35$^b$ & 2008-215 & 96  & 53--61 & 36--44   & 6.1--22.2  & ISS\_079RI\_RETARMRLP001\_PRIME \\
50$^b$ & 2008-237 & 95  & 51--59 & 37--46   & 5.9--17.8  & ISS\_082RI\_RETARMRLP001\_PRIME \\
36$^b$ & 2008-312 & 94  & 38--48 & 47--56   & 5.6--10.7  & ISS\_092RI\_RETARMRLP001\_PRIME\\
37$^b$ & 2008-343 & 94  & 20--36 & 58--74   & 4.1--6.3  & ISS\_096RI\_RETARMRMP001\_PRIME \\
53$^a$ & 2009-044 & 93  & 64--83 & 151--162 & 4.9--4.9  & ISS\_103RI\_SHRTMOV001\_PRIME \\
54$^a$ & 2009-056 & 93  & 75--88 & 159--162 & 4.9--4.9  & ISS\_104RI\_SHRTMOV002\_PRIME \\
80$^a$ & 2010-246 & 84  & 55--88 & 30--48   & 2.6--4.2  & ISS\_137RI\_BMOVIE001\_PRIME\\
  &  &   &  &    &   & ISS\_137RI\_BMOVIE002\_PRIME\\
  &  &   &  &    &   & ISS\_137RI\_BMOVIE003\_PRIME\\
83$^a$ & 2012-181 & 75  & 67--71 & 17--38   & 2.0--2.6  & ISS\_168RI\_MOONLETCD001\_PRIME \\
84$^a$ & 2012-182 & 75  & 87--87 & 26--26   & 6.0--6.0  & ISS\_168RB\_BMOVIE001\_PRIME \\
85$^a$ & 2012-205 & 75  & 67--70 & 51--77   & 2.0--2.5  & ISS\_169RI\_MOONLETCD001\_PRIME \\
87$^a$ & 2013-033 & 73  & 75--83 & 13--21   & 5.7--6.6  & ISS\_180RI\_BMOVIE001\_PRIME \\
 \end{tabular}

\caption{\small Details of data sets used in this study.  The incidence, emission, and phase angles are represented by $\Theta_0$, $\Theta$, and $g$, respectively.  $\Theta_0$ and $\Theta$ are referenced to the northern ring-plane normal.  Radial scale is km per pixel in the radial direction in the image.  Notes: (a) Ansa movie.  (b) Azimuthal scan. \label{tab:data}}

\end{table}
%--------------------------------------------------------------------

Prior work \cite{Porco1983} modeled the centerline of the ringlet by averaging models of the inner and outer edges, rather than examining the edges independently, because the early data sets typically had poor radial resolution and averaging of inner/outer edge measurements improves the signal to noise in the measurements.  Referring to Table \ref{tab:data}, the Cassini ISS data sets used in the present study have sufficient radial resolution to allow the ringlet edges to be examined independently, using the methods described in \cite{Spitale2009} (recall from that work that we measure edge radii with a precision of about 0.1 pixel).  Many of the data sets used in this study were also examined in the \cite{Spitale2010} B-ring study, and the labeling scheme is the same.  We focused on narrow-angle clear-filter images with no compression or lossless compression in which the ring, and the fiducial ring feature mentioned below, were visible and not significantly obscured by other objects like the planet or its shadow.

Data reduction is analogous to that in \cite{Spitale2009}: radiometric calibration was performed using CISSCAL (Cassini Imaging Science Subsystem Calibration)\cite{Porco2004}; pointing corrections were performed by aligning quasi-circular feature 13 (a ring edge in the Cassini division) from \cite{French1993} with its position in the images; a 5-pixel-wide radial profile was extracted from each image along a path with the best radial resolution; features in the profile were measured using the well-tested half-power method, as in \cite{Spitale2009}; final feature radii were computed relative to the measured radius of \cite{French1993} feature 13, though in this work we used updated parameters for that feature from \cite{Hedman2010}, who refer to that feature as the outer edge of the Russell gap and confirm its circularity (to 0.9 km).  The \cite{Hedman2010} radius is $\sim$1 km smaller and the standard deviation is $\sim$30\% larger than for the \cite{French1993} solution, but the two solutions are statistically consistent with one another.  Radial uncertainties were taken to be 0.1 times the observed width of the edge in the image, combined with the 0.9-km uncertainty in the \cite{Hedman2010} reference feature parameters.

%=======================================================================
\section{Keplerian Shape}  				\label{sec:m=1}
%=======================================================================

%--------------------------------------------------------------
% fit elements
%--------------------------------------------------------------
\begin{table}[h!]
 \tiny
 \center
 \begin{tabular}{lllllll}
 Fit:                           & 610           &               & 611           &               & 612           &               \\			      
 Element 			& Inner 	& Outer 	& Inner 	& Outer 	& Inner 	& Outer 	\\ \hline \hline
 $a$(km)			& 117803.24(7)  & 117824.00(7)  & 117802.14(6)  & 117823.06(7)  & 117803.16(5)  & 117824.08(6)  \\
 $e \times 10^{-4}$		& 2.390(8)      & 2.446(8)      & 2.407(9)      & 2.460(9)      & 2.392(8)      & 2.444(8)       \\
 $ae$(km)			& 28.157(3)     & 28.815(3)     & 28.356(4)     & 28.979(4)     & 28.182(3)     & 28.791(3)     \\
 $\varpi_0$($^\circ$)		& 333.6(2)      & 334.6(2)      & 333.1(2)      & 334.0(2)      & 333.5(2)      & 334.7(2)      \\
 $\Omega_p$($^\circ$/day)	& 5.0266(2)     & 5.0258(2)     & [5.02092]     & [5.02092]   & [5.0262]      & [5.0262]     \\ \hline

 statistics:			&               &               &               &               &               &               \\ \hline
 $\chi^2$/DOF			& 2.7           & 3.1 	        & 3.4           & 3.6           & 2.7	        & 3.6           \\
 RMS(km)			& 3.2           & 3.2 	        & 3.4           & 3.6           & 3.2	        & 3.6           \\ \hline

 \end{tabular}

\caption{\small Best-fit streamline elements for $m=1$ models of inner and outer edges of the Huygens ringlet for models with the apsidal precession rate $\Omega_p$ free (fit 610), and fixed to the average of the expected rates for the inner and outer edges (fit 611).  Longitudes are given at the central epoch, JED 2454935.8.  Quantities in brackets were held fixed.\label{tab:elem_m=1}}

\end{table}
%--------------------------------------------------------------

%--------------------------------------------------------------------
\begin{figure}[h!]
 \begin{center}
{\scalebox{0.7}{\includegraphics{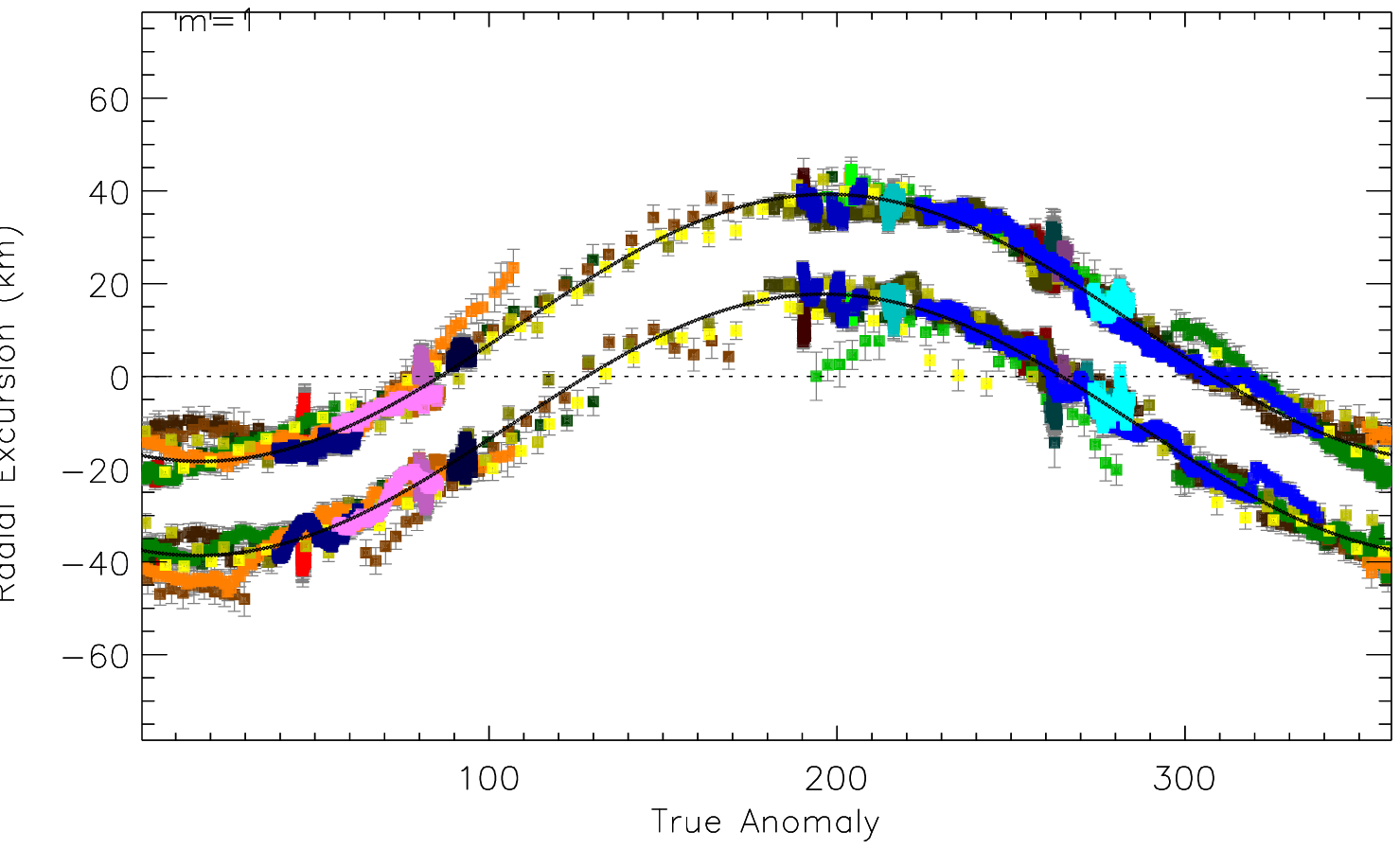}}}
{\scalebox{0.39}{\includegraphics{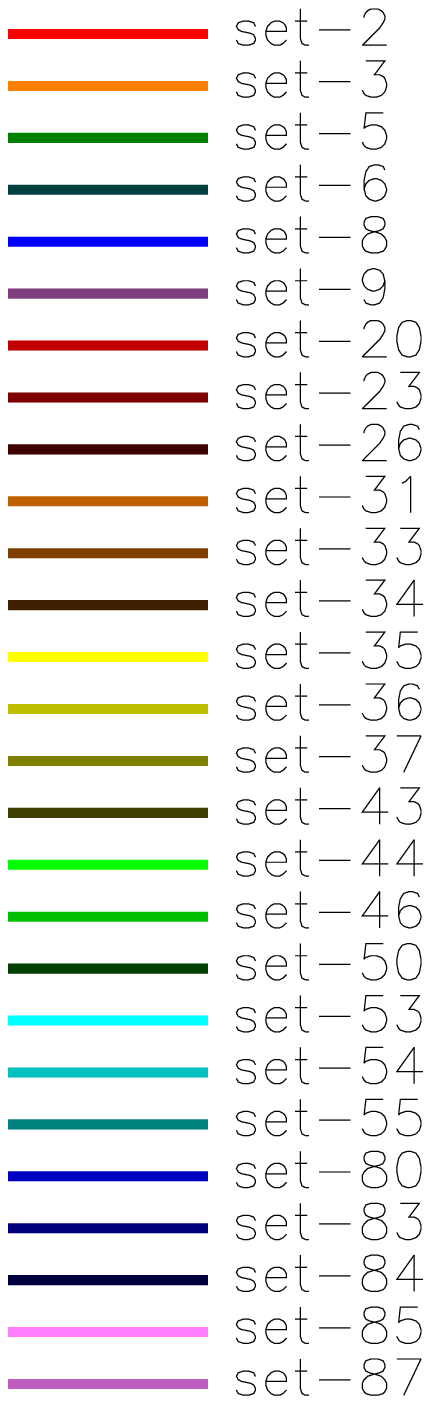}}}
 \caption{\small Radial excursion vs. true anomaly for the pure Keplerian model fits given in Table \ref{tab:elem_m=1} for fit 612.  Radial excursions are plotted relative to the centerline of the ringlet.  The true anomaly is $\lambda - \varpi$, where $\lambda$ is the measured longitude of the data point and the instantaneous orientation of the pattern is $\varpi = \varpi_0 + \Omega_p t$.  The model is shown with a solid curve.  Colors denote data sets (See Table \ref{tab:data}) according to the legend at right.}
 \label{fig:m=1}
\end{center}
\end{figure}
%--------------------------------------------------------------------

\cite{Porco1983} showed that the primary shape of the Huygens ringlet is a Keplerian ellipse (i.e., $m=1$ normal mode) with a radial amplitude of $\sim$ 32 km.  Therefore we begin by modeling a wave-number-1 pattern.  The first column in Table \ref{tab:elem_m=1} shows fits to the inner and outer edges of an uninclined ringlet with all parameters ($a$, $e$, $\varpi^{(m)}_0$, and $\Omega_p$) free (fit 610).  The amplitudes $ae$ are similar to, though somewhat smaller than that seen by \cite{Porco1983}.  The eccentricities of the edges are nearly the same, and at the central epoch of the fit the apses were nearly aligned, differing by about 1$^\circ$.  The measured pattern speeds $\Omega_p$ differ from one another by 4 standard deviations and are faster than expected based on Saturn's gravity field (5.02092$^\circ$/day) by more than 25 standard deviations.

In order to maintain apsidal alignment (from a geometric standpoint) the pattern speeds for the inner and outer edges should be the same over the long term.  The apsidal shifts (outer - inner) at the initial and final epochs implied by fit 610 are $\sim$2.1$^\circ$ and $\sim$-0.2$^\circ$ respectively, in both cases much less than would be necessary to destroy the pattern, so the Keplerian shape of the ringlet is coherent throughout the interval of the fit.  Given the relatively large RMS residuals in the fits, the difference in pattern speeds may not be meaningful anyway.  Therefore, we performed two additional fits with those speeds fixed to the same assumed central rate (Table \ref{tab:elem_m=1}): for fit 611, the pattern speeds are fixed to the expected rate at the center of the ringlet; for fit 612, they are fixed to the mean of the rates measured in fit 610.  Fit 611 (with the pattern speeds fixed at the expected rate) gives the poorest fit, with generally larger $\chi^2$ and RMS residuals than for fits 610 and 612.  Fit 612 (with the pattern speeds fixed at the mean measured rate) has similar statistics to fit 610.  Because there is no reason to impose the condition that the ringlet must precess at the expected rate, we take fit 612 as the most plausible fit to the $m=1$-only patterns.  Fig. \ref{fig:m=1} shows all of the data sets plotted using the parameters of fit 612.  The ringlet shape is largely described by the $m=1$ model, but other structure is apparent, indicated by the $\sim$3-km RMS residual, and as noted by \cite{Porco1983}.

%============================================================================
\section{Mode Searches}  				\label{sec:modes}
%============================================================================
The residuals to the $m=1$ fits in Table \ref{tab:elem_m=1} suggest that there may be other normal modes present in the ringlet edges.  Indeed, we have already noted that one or more forced wave-number-2 disturbances are likely to exist.  Therefore we performed a search for additional normal modes on each edge of the ringlet.  As in \cite{Spitale2010}, we modeled each ring edge as a linear combination of sinusoidal components, each precessing at its own rate.  To search for a normal mode with wave number $m$, we performed a suite of fits.  In each fit, the pattern speed $\Omega_p^{(m)}$ was fixed, while the other parameters -- $a$, $e$, $\varpi_0$, $\dot\varpi$, $e_m$, $\varpi_0^{(m)}$ -- were optimized.  Each search spanned a range of pattern speeds.  To evaluate the results, we plotted the mode amplitude, $A = a e_m$, divided by the RMS residual for the fit, as a function of the fixed pattern speed.  We label this quantity, $S \equiv A/RMS$, the "significance" of the fit.  For a given pattern speed, $S$ measures the strength of the mode relative to the variations in the pattern that are not modeled.  Those additional variations may be noise, other unknown normal modes, or some other irregular structure.  Unambiguous solutions should have a significance of unity or better, indicating that they are easily distinguished from the unmodeled variations.  A significance value between 0.5 and 1 suggests that a mode is present, but is difficult to discern, possibly because measurement errors are too large.  Such solutions may be worth considering if they are more significant than other solutions, or if their speeds are physically plausible.  Solutions with $S<0.5$ are questionable.  

We initially performed broad, unguided mode searches with wave numbers ranging from 0 to 50, and pattern speeds from about 300$^\circ$/day to 3000$^\circ$/day.  Few ISS data sets have more than $\sim$100 images required to formally resolve wave numbers higher than 50 (assuming uniform azimuthal coverage), and the range of pattern speeds corresponds to inner and outer Lindblad resonances located anywhere in the main rings.  However, due to the sparse and irregular sampling of the data sets, and the radial resolution of the imaging data sets, there are many local maxima, mostly with $S<0.5$.  Therefore, we performed targeted searches near pattern speeds with potential physical significance.  

For the forced modes, we focused on the parameter space near the Lindblad speeds appropriate for each wave number.  Pattern speeds for the free modes are more difficult to predict because they are not well understood.  The free normal modes detected on the outer edge of the B ring \cite{Spitale2010} had pattern speeds corresponding to Lindblad resonances somewhat interior to the edge.  In that work, the interpretation of the observed speeds was that the patterns represented density waves trapped in a resonant cavity bounded by the sharp ring edge and the Lindblad location for each given mode, which would select and amplify specific modes.  Those waves would propagate outward from the Lindblad location as trailing spirals.  That hypothesis has not been thoroughly investigated, but it was consistent with the observations in \cite{Spitale2010}, and  \cite{HahnSpitale2013} did not contradict it.  Another hypothesis is that the free modes are remnants of impulsive events.  \cite{HahnSpitale2013} showed that, once started, such modes may persist for decades or centuries, but that hypothesis did not explain the observed speeds of the free modes.  

Because the true origin of the free modes is unknown, in order to narrow down the range of pattern speeds to search, we followed the resonant cavity hypothesis because it predicts speeds for the free modes.  Accordingly, we searched for free modes on the outer edge with pattern speeds near those appropriate for Lindblad resonances in the Huygens ringlet, as was done for the force modes.  On the inner edge of the ringlet, any free normal modes would correspond to leading waves propagating \emph{inward} from \emph{outer} Lindblad resonances located in the interior of the ringlet.  Therefore, we additionally searched speeds near outer Lindblad resonances in the Huygens ringlet.  Those speeds are given by:
\begin{equation}
	m\Omega_p  = (m \mp 1) \pm \dot\varpi, %
                                                              \label{eq:speeds}
\end{equation}
where the upper sign refers to inner resonances, and the lower sign refers to outer resonances.

%--------------------------------------------------------------------
\begin{figure}[h!]
 \begin{center}
{\scalebox{0.4}{\includegraphics{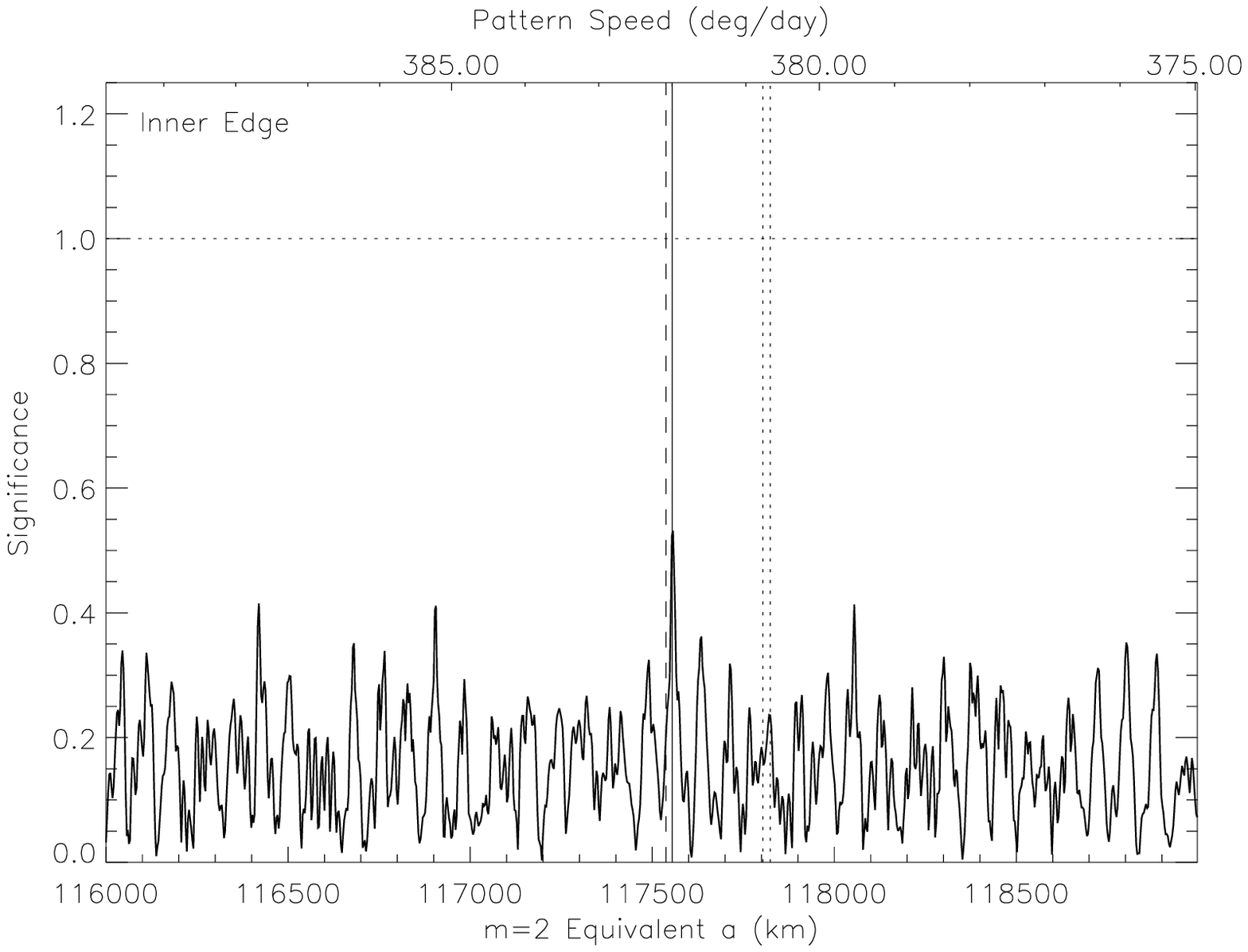}}}
{\scalebox{0.4}{\includegraphics{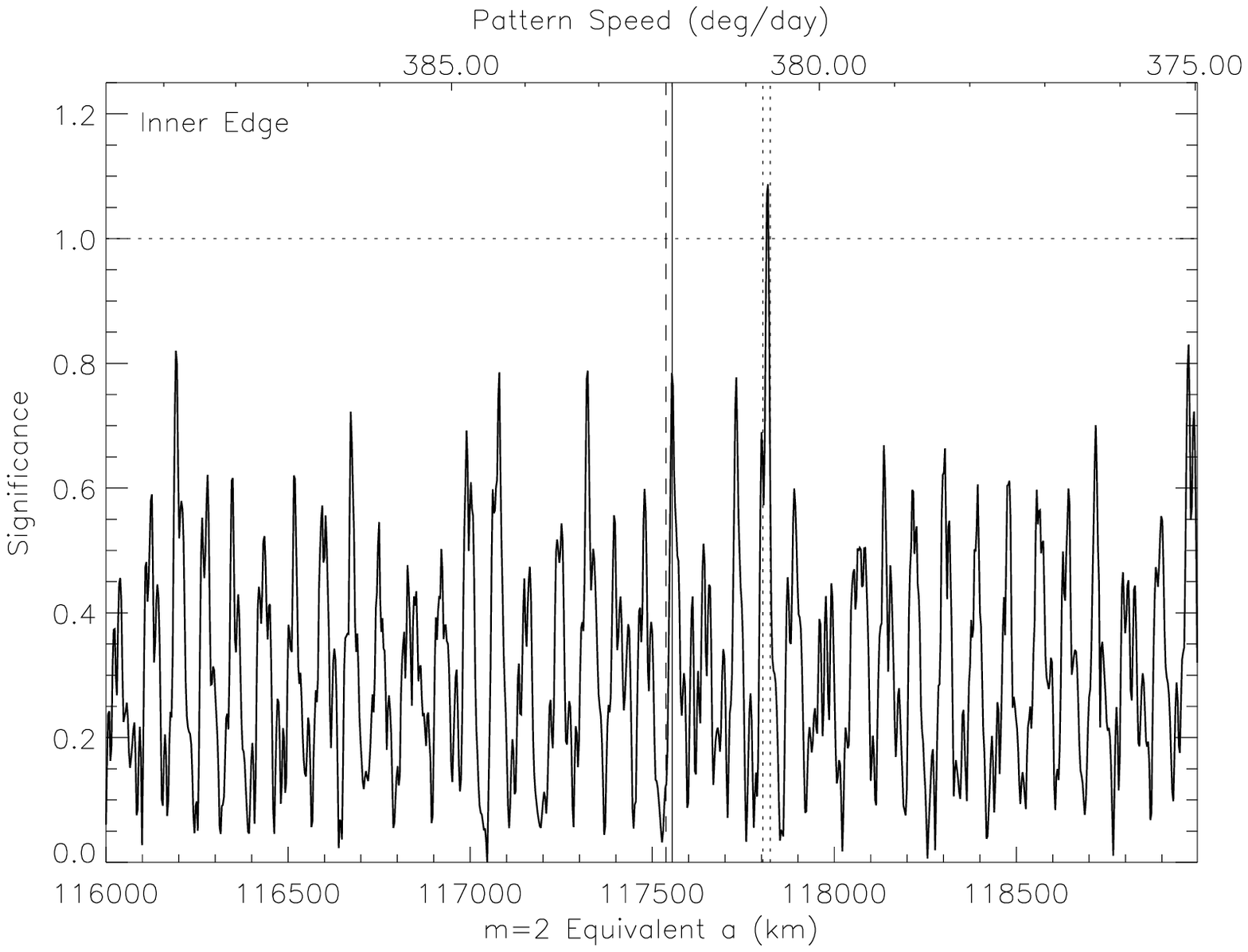}}}
 \caption{\small Parameter searches for $m=2$ modes on the inner and outer edges of the Huygens ringlet.  Plots show "Significance" $= A/RMS$ for a range of $m=2$ pattern speeds as a function of pattern speed and equivalent radial location (i.e., the location at which an m=2 pattern would precess at that rate).  For each pattern speed on the ordinate axis, all other ring orbital parameters are allowed to vary to find the best fit.  The solid vertical line is at Mimas' corrected mean motion (ee Sec. \ref{sec:m=2}), the dashed vertical line is at the speed of the B-ring free $m=2$ mode from \cite{Spitale2010}, and the dotted vertical lines are at the Lindblad speeds $m=2$ modes on the inner and outer edges of the Huygens ringlet.}
 \label{fig:span_m=2}
% fit 702
\end{center}
\end{figure}
%--------------------------------------------------------------------

Our unguided searches found no compelling evidence for normal modes with wave numbers besides 1 and 2.  The $m=1$ solution was already determined above (Table \ref{tab:elem_m=1}), so we performed $m=2$ searches relative to a model that accounted for the $m=1$ solutions specified by fit 612.  As shown in Fig. \ref{fig:span_m=2}, the $m=2$ solutions with pattern speeds at the expected forced and free rates were either the most significant or nearly so, though only one case -- the free $m=2$ mode on the outer edge -- showed a significance greater than 1.  Given the physical basis for their existence, we focused on those modes for more detailed investigation below (Sec. \ref{sec:m=2}).  No evidence was found for free normal modes on the inner edge of the ringlet, either near the expected pattern speed for an inward-propagating $m=2$ wave, or near speeds searched for the outer edge (that range of pattern speeds is close enough to Mimas' speed, that it was automatically covered in the search for a forced mode).

%--------------------------------------------------------------------
\begin{figure}[h!]
 \begin{center}
{\scalebox{0.4}{\includegraphics{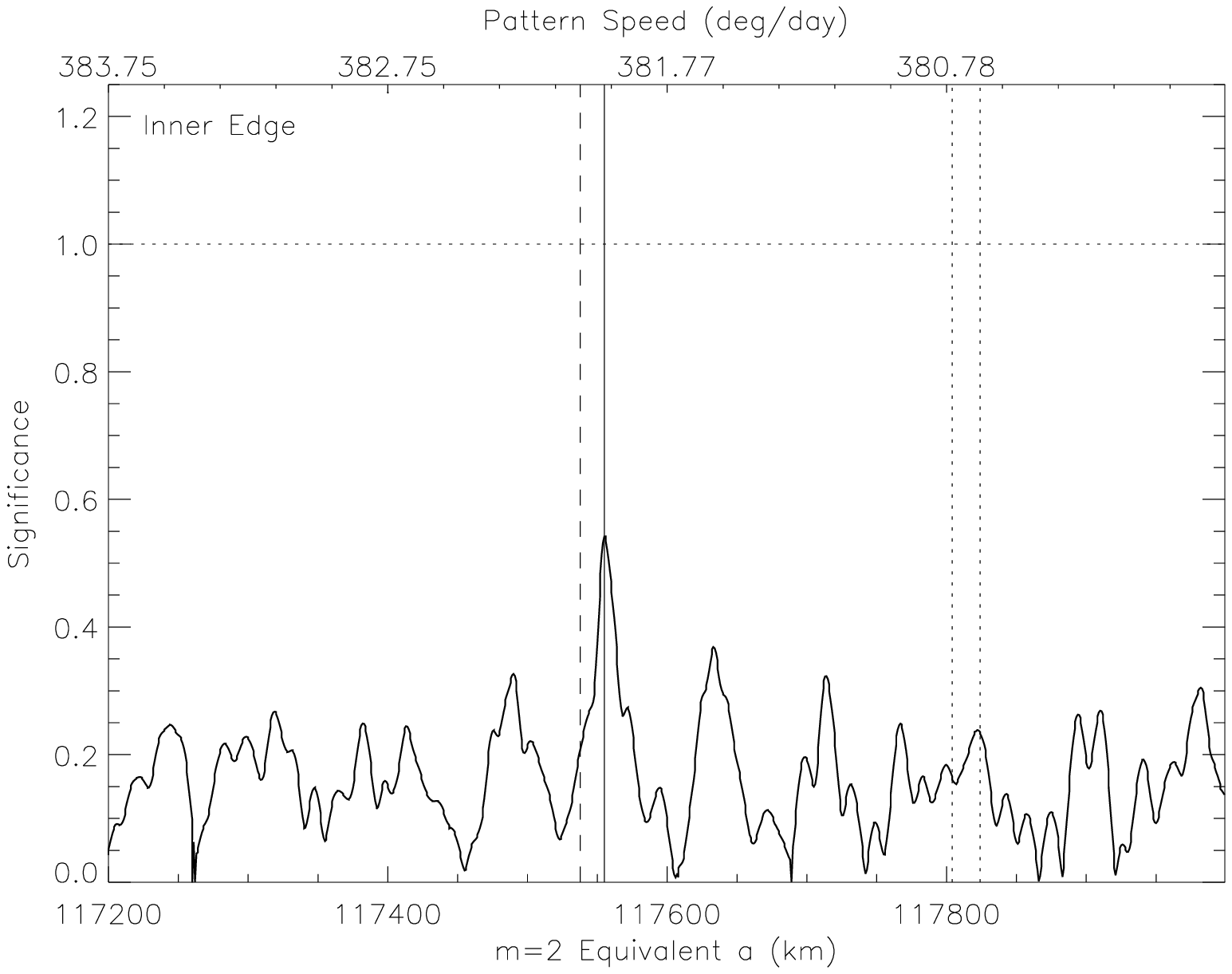}}}
{\scalebox{0.4}{\includegraphics{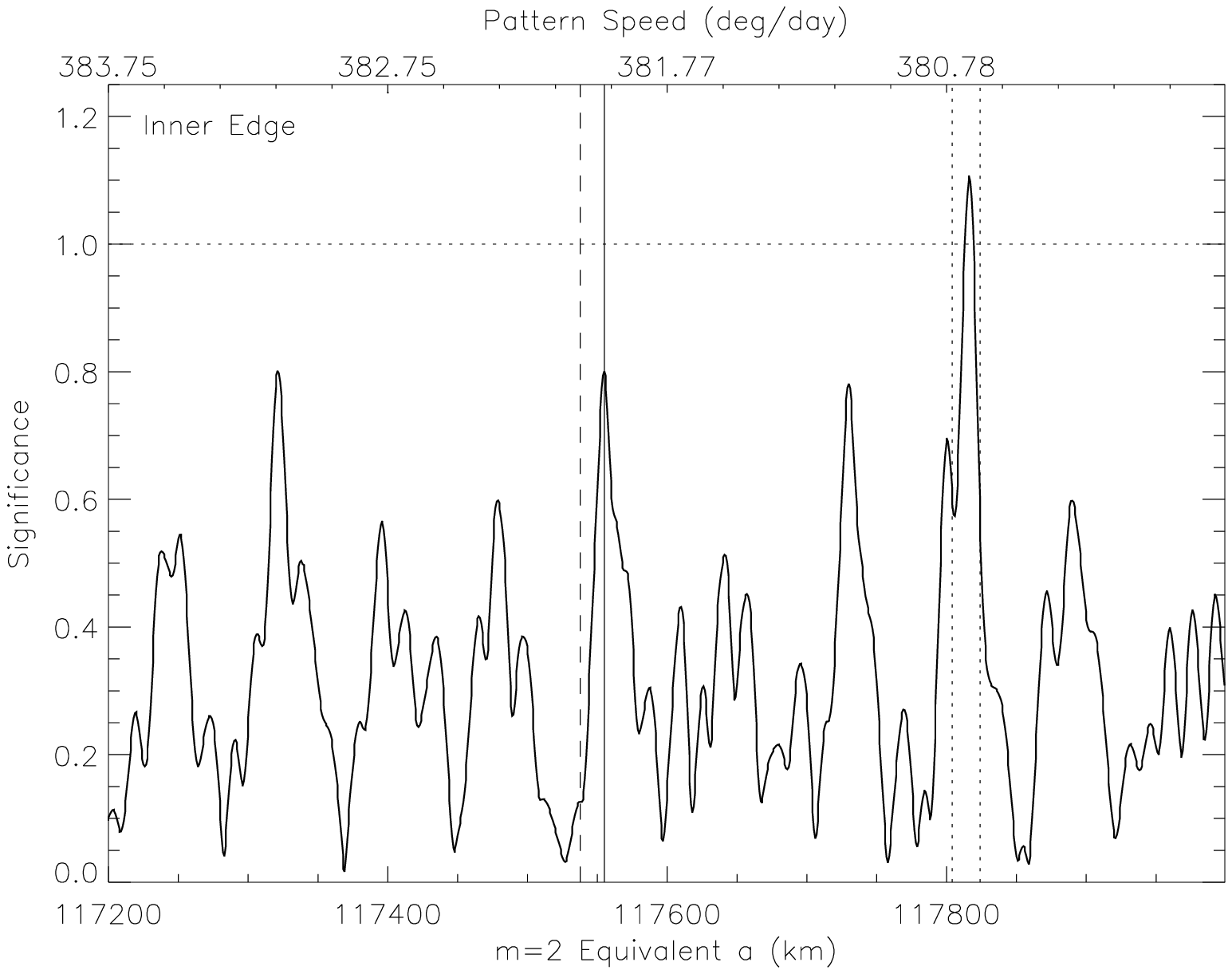}}}
 \caption{\small $m=2$ parameter searches from Fig. \ref{fig:span_m=2}, magnified to show detail in the vicinity of the Huygens ringlet.  The solid vertical line is at Mimas' corrected mean motion (see Sec. \ref{sec:m=2}), the dashed vertical line is at the speed of the B-ring free $m=2$ mode from \cite{Spitale2010}, and the dotted vertical lines are at the Lindblad speeds for $m=2$ modes on the inner and outer edges of the Huygens ringlet.}
 \label{fig:span_zoom_m=2}
% fit 702
\end{center}
\end{figure}
%--------------------------------------------------------------------

Fig. \ref{fig:span_zoom_m=2} shows the $m=2$ search results with a narrowed radial range.  On the inner edge, the most significant $m=2$ solution has $S \sim$ 0.55 and moves close to Mimas' speed.  The best solution on the outer edge has a significance of well over unity and has a pattern speed near that expected for an $m=2$ Lindblad resonance near the outer edge of the ringlet.  The outer edge also shows a solution near Mimas' speed with $S \sim$ 0.8, but there are other nearby solutions with comparable significance values, so further evidence will be required to accept that as a real solution (see below).  

We performed another search for wave numbers larger than 2, with the $m=2$ modes seen in the initial search above added to the kinematic model.  With the updated models, any additional normal modes should have higher significance values than in the initial searches because the previously unmodeled $m=2$ modes have been accounted for.  The results of the new searches were dominated by aliasing and still did not provide conclusive evidence of real normal modes.  

From this analysis, we conclude that there is an $m=2$ mode forced by Mimas on the inner edge of the ringlet, and an $m=2$ free mode on the outer edge.  There is probably a mode forced by Mimas on the outer edge as well.  It is possible that free modes moving at speeds significantly different than those that we focused on might have been missed in the unguided search.  Other modes likely do exist (e.g., \cite{Nicholson2014}), but their amplitudes are too small to detect in the ISS data sets.  The $m=2$ solutions found here are examined more closely in Sec. \ref{sec:m=2}.

%=======================================================================
\section{$m=2$ Solutions}  				\label{sec:m=2}
%=======================================================================
Motivated by the results of the mode search, we performed three-component fits to the ringlet's inner and outer edges where we included the $m=1$ Keplerian shape and two $m=2$ modes, one forced and one free.  Although no evidence for a free $m=2$ mode was seen on the inner edge, we included such a mode, but zeroed the amplitude for consistency in the fitting and plotting procedures.  In all fits, the inner and outer $m=1$ pattern speeds were assumed to be the same and were fixed to the value used in fit 612, i.e., the mean of the inner and outer rates determined in fit 610; other $m=1$ parameters were allowed to vary.  The forced $m=2$ part of the fit is complicated by the fact that Mimas' mean motion varies due to its resonance with Tethys with a period of $\sim$71 yr.  It is not known on what timescale the ring's response follows that variation in speed, so we performed fits with the forced $m=2$ pattern speed fixed to both Mimas' long-term average mean motion (henceforth referred to as its long-term mean rate) and its average mean motion during the interval spanned by the data sets (henceforth referred to as its short-term mean rate).  Moreover, we performed fits with one $m=2$ apoapse locked to Mimas' mean longitude (as would be expected for the ideal case with no dissipation) and fits where the periapse longitude was allowed to vary.

%--------------------------------------------------------------------------
\begin{table}[h!]
 \tiny
 \center
% note Mimas aligned with on apoapse of inner forced pattern.
 \begin{tabular}{lllllllll}
 Fit:				& 744 (adopted fit) &  			& 745 &   			& 746 &   			& 747 &   \\ 				
 Element 			& Inner 	& Outer 	& Inner 	& Outer 	& Inner 	& Outer 	& Inner 	& Outer        \\ \hline \hline
 $m$=1:				&               & 		& 		& 		& 		& 		& 		&              \\ \hline
 $a$(km)			& 117803.04(5)  & 117824.67(5)  & 117803.06(5)  & 117824.63(5)  & 117803.22(5)  & 117824.75(5)  & 117803.20(5)  & 117824.72(5) \\
 $e \times 10^{-4}$		& 2.404(7)      & 2.421(6)      & 2.402(7)      & 2.429(6)      & 2.401(7)      & 2.417(6)      & 2.403(7)      & 2.428(6)     \\
 $ae$(km)			& 28.320(3)     & 28.521(2)     & 28.301(3)     & 28.624(2)     & 28.288(3)     & 28.480(2)     & 28.309(3)     & 28.603(2)    \\
 $\varpi_0$($^\circ$)		& 334.1(2)      & 334.2(1)      & 334.0(2)      & 334.3(1)      & 333.7(2)      & 334.0(1)      & 333.7(2)      & 334.1(1)     \\
 $\Omega_p$($^\circ$/day)	& [5.02619]     & [5.02619]     & [5.02619]     & [5.02619]     & [5.02619]     & [5.02619]     & [5.02619]     & [5.02619]    \\ \hline

 $m$=2 (forced):		&               &               &   	        &               &               &               &               &              \\ \hline
 $e \times 10^{-4}$		& 0.137(6)      & 0.102(6)      & 0.137(6)      & 0.077(6)      & 0.131(6)      & 0.081(6)      & 0.131(6)      & 0.052(6)     \\
 $ae$(km)			& 1.62(4)	& 1.20(6)	& 1.62(4)	& 0.90(7)	& 1.55(5)	& 0.95(7)	& 1.55(5)	& 0.6(1)	\\
 $\varpi^{(0)}_0$($^\circ$)	& 33(2)         & 52(3)         & [34.341026]   & [34.341026]   & 35(3)         & 57(4)         & [34.341026]   & [34.341026]  \\
 $\Omega^{(0)}_p$($^\circ$/day)	& [381.9842(1)]	& [381.9842(1)]	& [381.9842(1)]	& [381.9842(1)] & [381.9945(1)]	& [381.9945(1)]	& [381.9945(1)]	& [381.9945(1)]  \\ \hline

 $m$=2 (free):			&               &               &   	        &               &               &               &               &              \\ \hline
 $e \times 10^{-4}$		& [0]           & 0.189(5)      & [0]           &  0.206(5)     & [0]           & 0.195(6)      & [0]           & 0.217(5)     \\
 $ae$(km)			& [0]           & 2.22(3)       & [0]           &  2.43(3)      & [0]           & 2.30(3)       & [0]           & 2.55(2)      \\
 $\varpi^{(f)}_0$($^\circ$)	& [0]           & 166(2)        & [0]	        &  168(1)       & [0]           & 165(2)        & [0]           & 170(1)       \\
 $\Omega^{(f)}_p$($^\circ$/day)	& [1132.2217]   & 380.698(2)    & [1132.2217]   &  380.697(2)   & [1132.2217]   & 380.697(2)    & [1132.2217]   & 380.697(2)   \\ \hline

 statistics:			&               &               &               &               &               &               &               &              \\ \hline
 $\chi^2$/DOF			& 2.2           & 1.4           & 2.2           & 1.5           & 2.3           & 1.5           & 2.3           & 2.3          \\
 RMS(km)			& 3.0           & 2.4           & 3.0           & 2.4           & 3.0           & 2.4           & 3.0           & 3.0          \\ \hline

 \end{tabular}

\caption{\small Best-fit streamline elements for 3-component models of the Huygens ringlet inner and outer edges.  Longitudes are given at the central epoch, JED 2454935.8 (UTC 2009-104T06:09:29.465).  Quantities in brackets were held fixed.  At the central epoch, Mimas' true longitude, as derived from the Jet Propulsion Laboratory (JPL) Navigation and Ancillary Information Facility (NAIF) kernels, was 124.341026$^\circ$, which differs by up to $\sim$1$^\circ$ from its mean longitude.  Hence, one maximum of the pattern was located at 34$\pm$1$^\circ$. \label{tab:elem_m=122}}

\end{table}
%--------------------------------------------------------------------------

Our model of Mimas' longitude variation was provided by R. A. Jacobson (2013, private communication).  It describes Mimas' mean longitude rate as a sinusoid with a period of about 71 yr and a long-term mean of 381.994495$^\circ$/day.  Over the interval of our observations, the Jacobson model yields a short-term mean rate of 381.98423$^\circ$/day.  Jacobson also provided along-track residuals for those orbit fits in km from the predicted position.  The along-track residuals vary quasi-sinusoidally with a period of $P \sim$ 24 yr.  The RMS value of the variation is $\sigma_x \sim \pm$2100 km.  As the variation occurs over that range twice per "period," we estimate the rate uncertainty as:
\begin{equation}
	\sigma_{\Omega_p}  = \frac{\sigma_x}{a_s} \frac{2}{P}. %
                                                                  \label{eq:jac}
\end{equation}
Taking Mimas' semimajor axis $a_s$ as 185539.5 km, Eq. \ref{eq:jac} yields a rate uncertainty of $\sigma_{\Omega_p} \simeq$ 0.0001$^\circ$/day.  Hence, we took Mimas' long-term mean rate to be 381.9945(1)$^\circ$/day and its short-term mean rate to be 381.9842(1)$^\circ$/day.  

Table \ref{tab:elem_m=122} shows the results of the three-component fits.  Parameters for the $m=1$ modes are close to those determined in the $m=1$-only fits (fits 612, Table \ref{fig:m=1}).  The apses  of the inner and outer $m=1$ patterns differ by 0.4$^\circ$ or less.  Consistent with the results of the mode search (see Fig. \ref{fig:span_zoom_m=2}), the solutions for the inner edge primarily show the expected forced $m=2$ mode, and the outer edge solutions show a strong free mode with a smaller forced mode.  We attempted fits on the inner edge with the free mode parameters variable and the results were statistically equivalent to those presented, justifying zeroing out that mode in each fit.

%--------------------------------------------------------------------------
\begin{figure}[h!]
 \begin{center}
{\scalebox{0.7}{\includegraphics{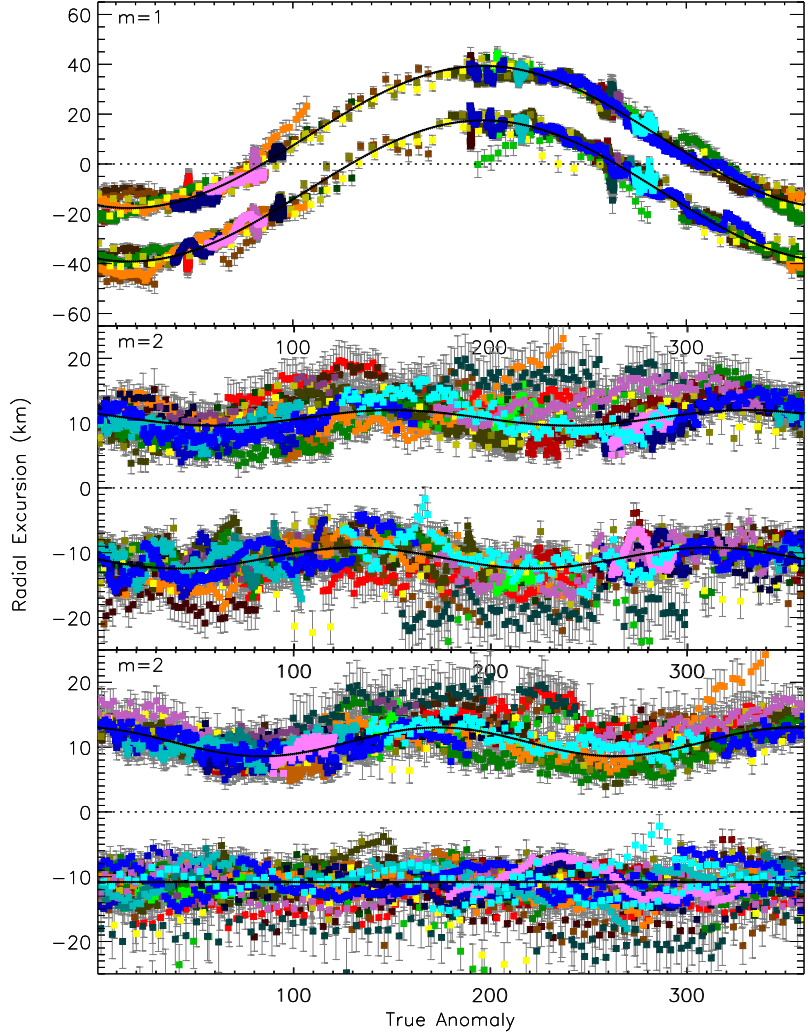}}}
 \caption{\small Radial excursion vs. true anomaly for each component of the three-component model fits given in Table \ref{tab:elem_m=122} for fit 744.  Radial excursions are plotted relative to the centerline of the ringlet.  The true anomaly is $\lambda - \varpi^{(m)}$, where $\lambda$ is the measured longitude of the data point and the instantaneous orientation of the pattern is $\varpi^{(m)} = \varpi^{(m)}_0 + \Omega_p t$.  In each panel, only the residuals from the panel above are plotted.  The model is shown with a solid curve.  Colors denote data sets according to the legend in Fig. \ref{fig:m=1}.}
 \label{fig:m=122}
\end{center}
\end{figure}
%--------------------------------------------------------------------------

The fit phases of the forced modes on the inner edge align one apoapse of the $m=2$ pattern within Mimas' longitude, whether the pattern speed is fixed at Mimas' short-term mean rate (fit 744), or fixed at Mimas' long-term mean rate (fit 746).  On the outer edge, the apoapses of the forced solutions lead Mimas' longitude by $\sim20^\circ$.  The fit amplitudes on the inner edge are identical whether or not the phase is fixed, whereas the amplitudes on the outer edge are considerably larger (with somewhat smaller formal uncertainties) when the phase is allowed to vary in the fits rather than being fixed to Mimas.  

Using the \cite{Goldreich1982} formulation for a free particle perturbed by a Lindblad resonance, we expect particles orbiting at semimajor axis $a$ to oscillate radially in response to the Mimas resonance with an amplitude $ae_M = |A/(a-a_M)|$, where:
\begin{equation}
	A = \frac{M_s}{M_p} \frac{\alpha a^2}{3 (m - 1)} %
	         \left.\left( 2m + \alpha\frac{d}{d\alpha} \right) %
	              b_{1/2}^{(m)}(\alpha)\right|_{a=a_M}, %
                                                                  \label{eq:bbb}
\end{equation}
$a_M$ is the resonance location, and $\alpha = a/a_s$.  Setting $a_s = 185539.5$ km, Mimas' orbital semimajor axis, and noting that the expression involving the Laplace coefficient $b_{1/2}^{(m)}(\alpha)$ evaluates to $\sim$ 2.38 at the inner edge of the ringlet (see Tables 8.1 and 8.5 of \cite{MurrayDermott}), the computed amplitude is $ae_M = $ 1.9 km, similar to the amplitudes of our forced solutions on the inner edge.  

The width of the ringlet is much smaller than its distance from the resonance, so Eq. \ref{eq:bbb} predicts about the same forced amplitude for the outer as for the inner edge.  However, the fit amplitudes for the forced mode on the outer edge are significantly smaller than that prediction for all four fits.  The largest amplitudes, and (marginally) smallest uncertainties, occur for the fits (744 and 746) where the orientation of the forced mode was not fixed to Mimas' longitude.  The $\sim20^\circ$ phase lag (these particles orbit faster than Mimas, so their \emph{leading} longitudes imply they \emph{lag} the perturbation in time) implies significant dissipation, which would explain the reduced amplitudes relative to the prediction for isolated particles.  Therefore, because fits 745 and 747, in which the phases of the forced modes on both edges were forced to track Mimas, do not explain why the forced $m=2$ amplitudes on the outer edge are so much smaller than predicted by the single-particle theory, they were ruled out.  As the statistics for fit 744 show a marginally better fit than fit 746, we chose fit 744 as the most likely model of the Huygens ringlet up to wave number 2.  Fig. \ref{fig:m=122} plots radial excursions vs. true anomaly for the each component of the adopted three-component model, on both the inner and outer edges.

Following \cite{Borderies1982}, the $\Delta \sim20^\circ$ phase lag can be related to a kinematic viscosity $\nu$ via:
\begin{equation}
    \sin m\Delta = \frac{\nu}{n(m a)^2}\left(\frac{M_p}{M_s}\right)^2, \\
							      \label{eq:nu}
\end{equation}
where $n$ is the mean motion.  The resulting value of $\nu \sim 240$ cm$^2$/s falls on the higher end of the various estimates for the kinematic viscosity in the A ring \cite{Tiscareno2007, Chakrabati1989, Shu1985, Lissauer1984, Esposito1983}.  The RMS random velocity is given by \cite{Goldreich1982}:
\begin{equation}
    v^2 \sim 2\nu n\left( \frac{1 + \tau^2}{\tau} \right). \\
							      \label{eq:v2}
\end{equation}
Assuming an optical thickness $\tau$ of 1.5 \cite{French1993}, Eq. \ref{eq:v2} yields $v \sim $ 4 mm/s.  The vertical scale height is then roughly $H \sim v/n \simeq 25$ m.  In contrast, it can be inferred from \cite{Nicholson2014} and \cite{French2015} that the Maxwell and Colombo ringlets have scale heights around 5 m or less.

%==================================================================
\section{Time Variability}    \label{sec:variability}
%==================================================================
Here we examine the shape of each edge of the ringlet on a per-observation basis.  Our ability to determine a meaningful $m=1$ shape at each epoch depends on the azimuthal sampling, which is satisfied for five of the azimuthal scans; the ansa movies, which sample a very narrow range of inertial longitudes, yield very limited true anomaly coverage for the $m=1$ pattern because of its slow precession rate.  However, because of the much faster speed of the $m=2$ patterns, the $m=2$ true anomaly coverage is sufficient to obtain an $m=2$ model in many cases where an $m=1$ model cannot be obtained.  In those cases, we fit a fictitious $m=1$ model (i.e., $m=1$ parameters were fit along with the $m=2$ paramters in order to model out the long-wavelength variation) with the understanding that only the $m=2$ component can be trusted.  Specifically, the $m=1$ patterns for sets 3, 5, and 43 capture only points near the ringlet periapse or apoapse, so the fit eccentricities are not reliable, though those fits do yield reliable instantaneous $m=2$ parameters with acceptable uncertainties.  The ansa movies capture virtually no $m=1$ variation, so in those cases an $m=2$-only model was fit, i.e., the $m=1$ eccentricity was set to zero.  For all of the fits, the $m=1$ precession rate was fixed to the value from in fit 612 (see Sec.\ref{sec:m=1}) because the $m=1$ orientation changes by at most a few degrees during the observation, giving the fitting procedure little leverage with which to constrain that parameter.  For $m=2$ component, we included only a single mode, even on the outer edge, because the short duration of the observations does not effectively constrain the pattern speeds of those modes, so a fit with two modes with similar pattern speeds would produce an ambiguous result.  Instead, we fixed the $m=2$ pattern speed to Mimas' long-term mean rate; we could have chosen other similar rates, but the differences in the resulting pattern orientations are insignificant over these short timescales.  Any effect of multiple $m=2$ pattern will show up as a beat pattern.

%--------------------------------------------------------------------------
\begin{figure}[h!]
 \begin{center}
{\scalebox{0.5}{\includegraphics{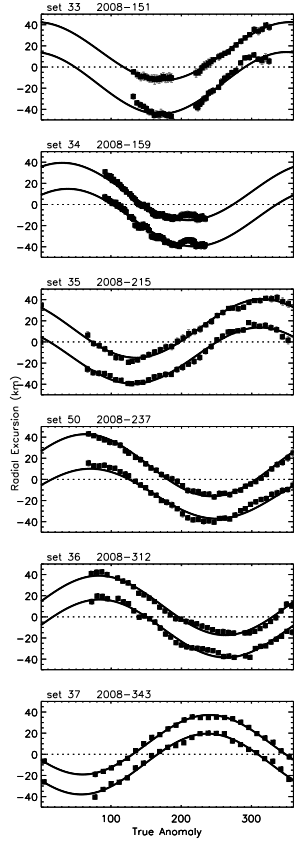}}}
 \caption{\small Radial excursion vs. true anomaly for the Keplerian components of fits to individual data sets where the Keplerian mode can be determined from that data set alone.  Radial excursions are plotted relative to the centerline of the ringlet.  The true anomaly is $\lambda - \varpi$, where $\lambda$ is the measured longitude of the data point and the instantaneous orientation of the pattern is $\varpi = \varpi_0 + \Omega_p t$.  The model is shown with a solid curve.}
 \label{fig:m=1_each}
% fits 1010...
\end{center}
\end{figure}
%--------------------------------------------------------------------------

%--------------------------------------------------------------------------
\begin{figure}[h!]
 \begin{center}
{\scalebox{0.5}{\includegraphics{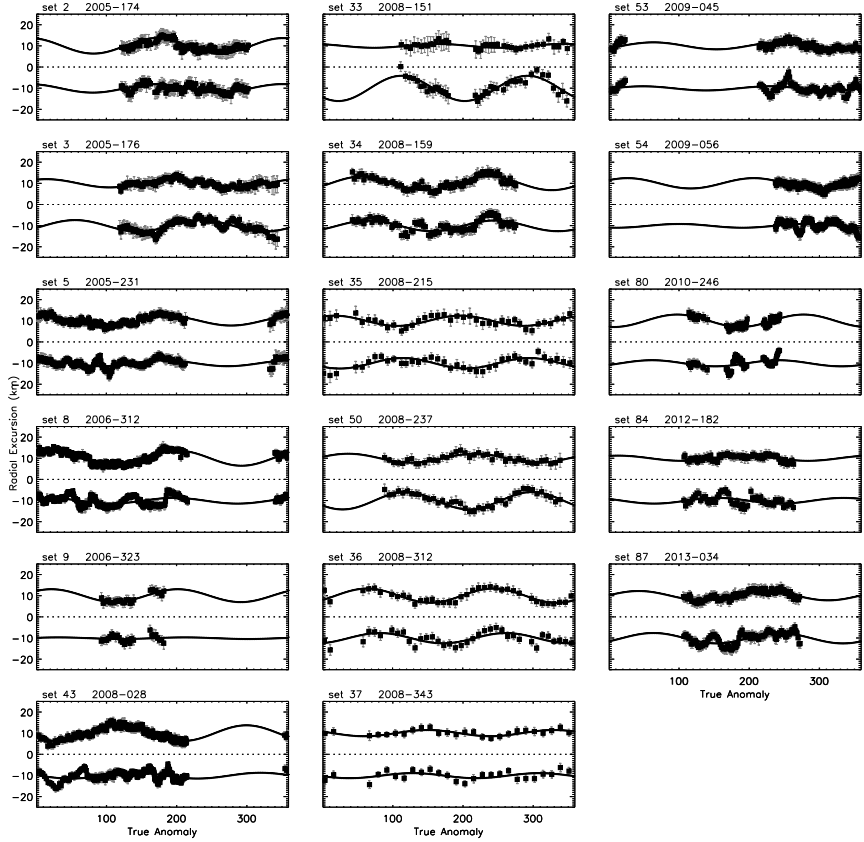}}}
 \caption{\small Radial excursion vs. true anomaly for the total $m=2$ components of fits to individual data sets where the $m=2$ mode can be determined from that data set alone.  Radial excursions are plotted relative to the centerline of the ringlet.  The true anomaly is $\lambda - \varpi$, where $\lambda$ is the measured longitude of the data point and the instantaneous orientation of the pattern is $\varpi = \varpi_0 + \Omega_p t$.  The model is shown with a solid curve.}
 \label{fig:m=2_each}
% fits 1001...
\end{center}
\end{figure}
%--------------------------------------------------------------------------

Figs. \ref{fig:m=1_each} and \ref{fig:m=2_each} plot the results of these fits.  We display only the results that we consider to be reliable based on the considerations in the above paragraph.  For the $m=2$ fits (Fig. \ref{fig:m=2_each}) where a fictitious $m=1$ pattern was used, the apparent mean width of the ringlet varied significantly depending on the radial differences in the $m=1$ patterns at the longitudes where the points occurred.  Because that width variation is not real, we adjusted the mean radii of the patterns so that the mean widths seen in the plots corresponds to the mean width of the ringlet from fit 612.  Quantitative fit parameters and statistics are not provided as these fits are intended only to facilitate a qualitative discussion.

%--------------------------------------------------------------------------
\begin{figure}[h!]
 \begin{center}
{\scalebox{0.7}{\includegraphics{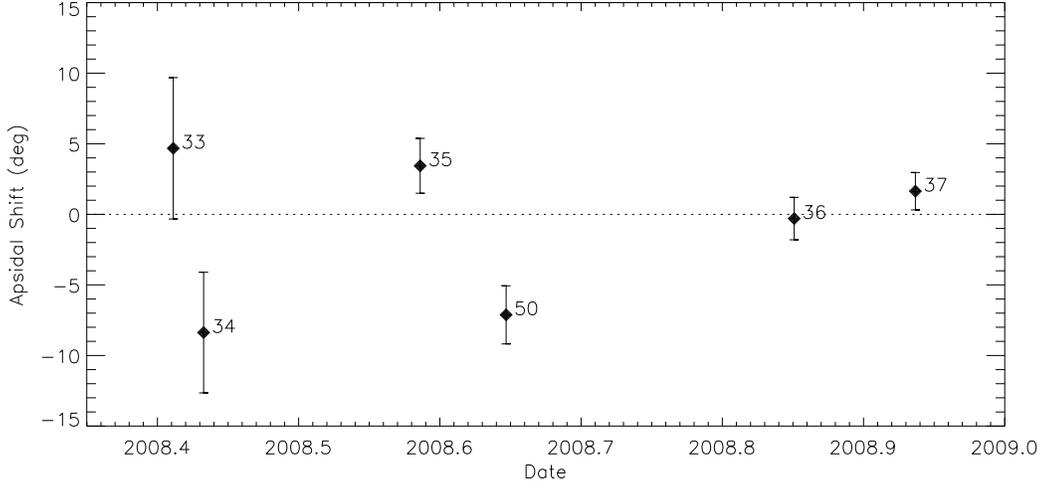}}}
 \caption{\small Apsidal shifts ($\varpi_0^{\mathrm{(outer)}}-\varpi_0^{\mathrm{(inner)}}$) for the data sets for which reliable periapse longitudes could be obtained.}
 \label{fig:dlp}
\end{center}
\end{figure}
%--------------------------------------------------------------------------

Fig. \ref{fig:dlp} shows the apsidal shifts for the data sets plotted in Fig. \ref{fig:m=1_each}.  The pattern of apsidal misalignments is suggestive of libration about a state of apse alignment during that $\sim7$-mo period in 2008.  However, there is not enough information to fit a period, or to establish unambiguously that the variation is even periodic, though that would be the simplest interpretation.

%--------------------------------------------------------------------------
\begin{figure}[h!]
 \begin{center}
{\scalebox{0.7}{\includegraphics{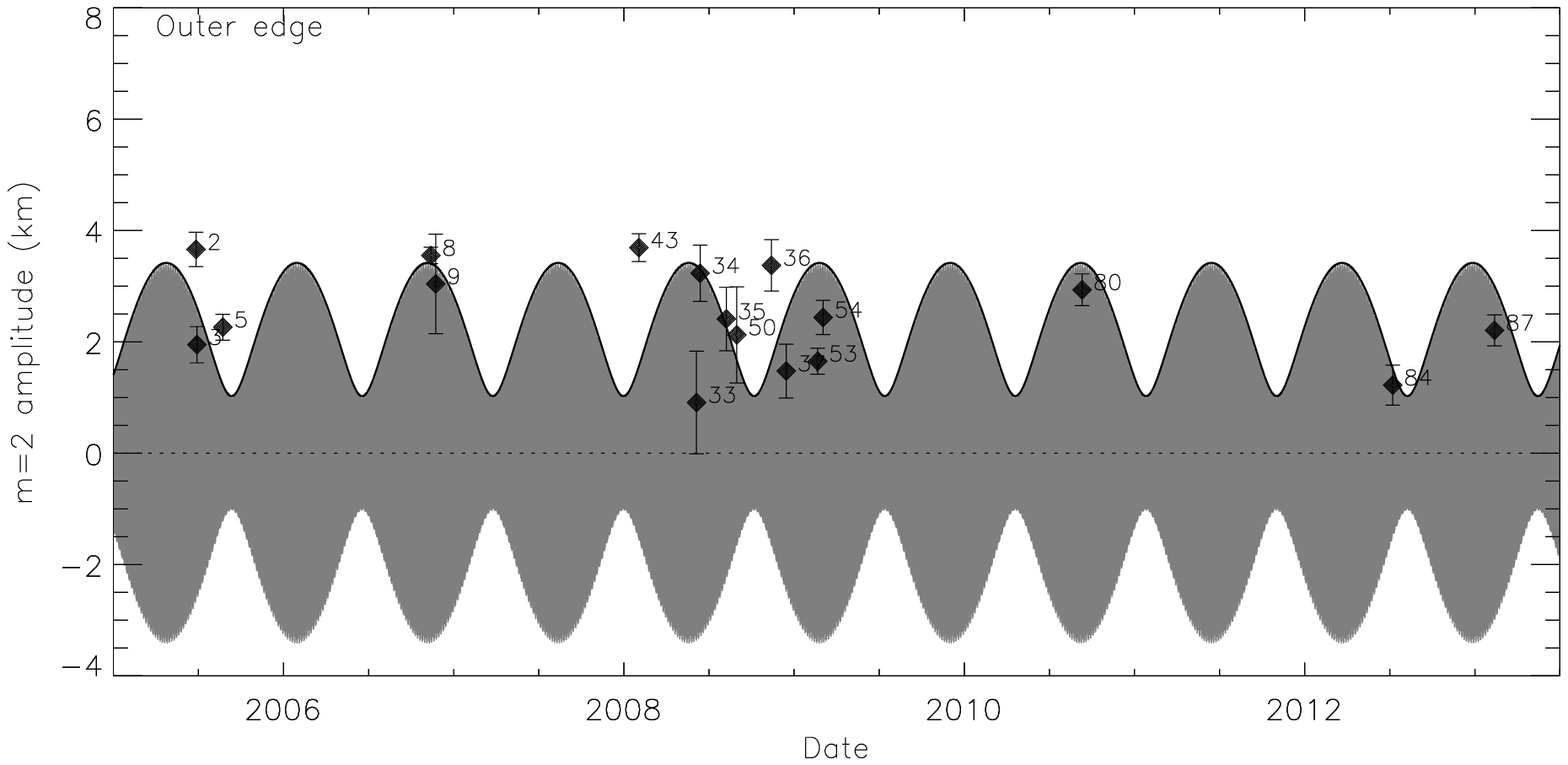}}}
{\scalebox{0.7}{\includegraphics{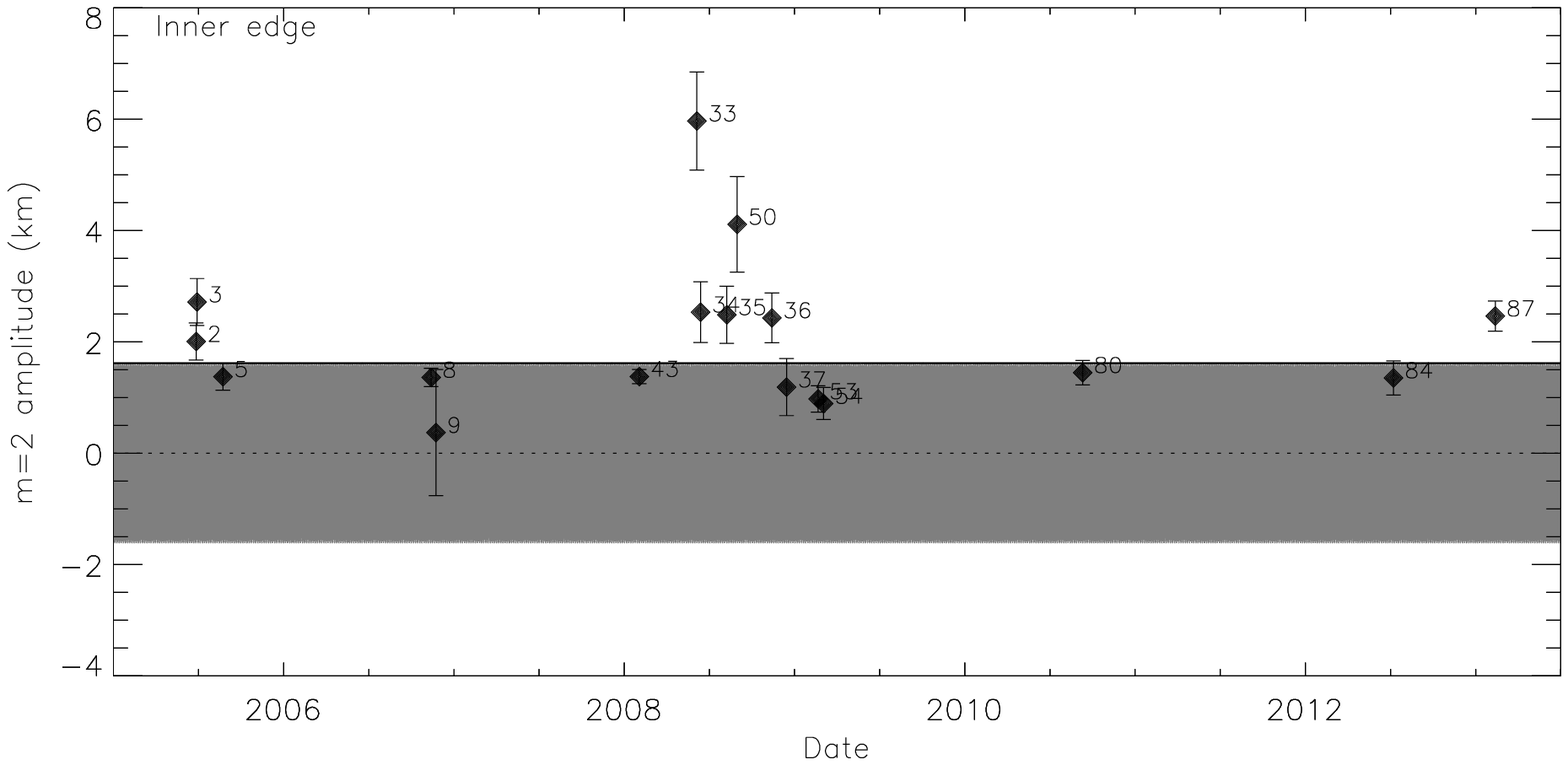}}}
 \caption{\small Amplitudes $ae_2$ of the total $m=2$ pattern for each data set plotted in Fig. \ref{fig:m=2_each}.  The shaded gray area consists of finely spaced gray curves giving the instantaneous radial excursion from the Keplerian model of fit 744. The solid line shows the beat envelope for the $m=2$ components of that fit.}
 \label{fig:amplitudes}
\end{center}
\end{figure}
%--------------------------------------------------------------------------

Fig. \ref{fig:amplitudes} shows the total $m=2$ amplitudes plotted as a function of time.  Also plotted are (gray) the instantaneous radial excursions from the pure Keplerian component of fit 744 and (solid) the beat envelope (see Eq. 11 of \cite{Spitale2010}) for the $m=2$ modes from fit 744, which is flat for the inner edge because there is only a single $m=2$ mode.

If there are no other perturbations on the ringlet besides those two $m=2$ patterns, then the total $m=2$ amplitudes should fall on the beat curve.  On the outer edge, many of the amplitudes do fall near the beat curve, and those that do not are still generally consistent with the expected overall variation in $m=2$ amplitude.  On the inner edge, the amplitudes are generally scattered near the (flat) curve, except for sets 50 and 33, which have amplitudes roughly double and triple those expected from the global fit, respectively.  The cause of these deviations is likely the presence of additional systematic structure that is not modeled in the three-component fit, discussed in the next section.  In some data sets, such structures may masquerade as portions of an $m=2$ pattern, resulting spurious large fit amplitudes.

%==================================================================
\section{Irregular Structure}  \label{sec:irregular}
%==================================================================
Thus far we have quantified all of the forced and free normal modes that are evident at the inner and outer edges of the Huygens ringlet.  Nonetheless, Figs. \ref{fig:m=122} and \ref{fig:amplitudes} demonstrate that there are significant discrepancies between the observed shapes and and the kinematic models of those edges; we refer to those discrepancies as "irregular structure."  In \cite{Spitale2010}, we identified at least 2 non-periodic perturbations (referred to as "spikes") in the B-ring edge that moved at Keplerian rates, suggesting the presence of mass concentrations internal to the ring, close enough to the edge to produce observable radial anomalies.  A similar disturbance had also been noted at the A-ring edge \cite{Spitale2009}, and more recently an object has been directly imaged on that edge \cite{Murray2014}.  The irregular appearance of the Huygens ringlet suggests the presence of many such mass inhomogeneities.

%--------------------------------------------------------------------------
\begin{figure}[h!]
 \begin{center}

{\scalebox{0.4}{\includegraphics{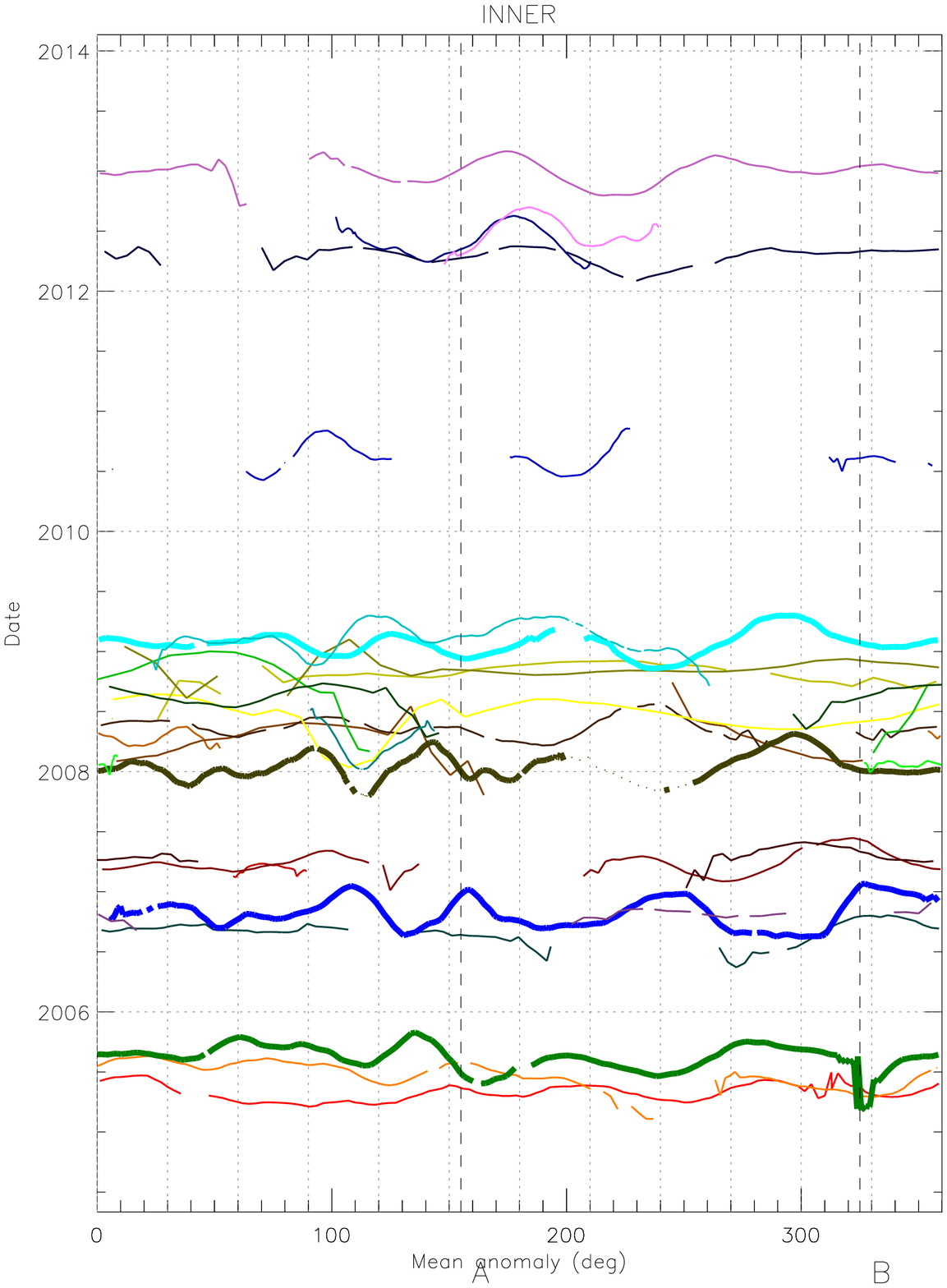}}}
{\scalebox{0.4}{\includegraphics{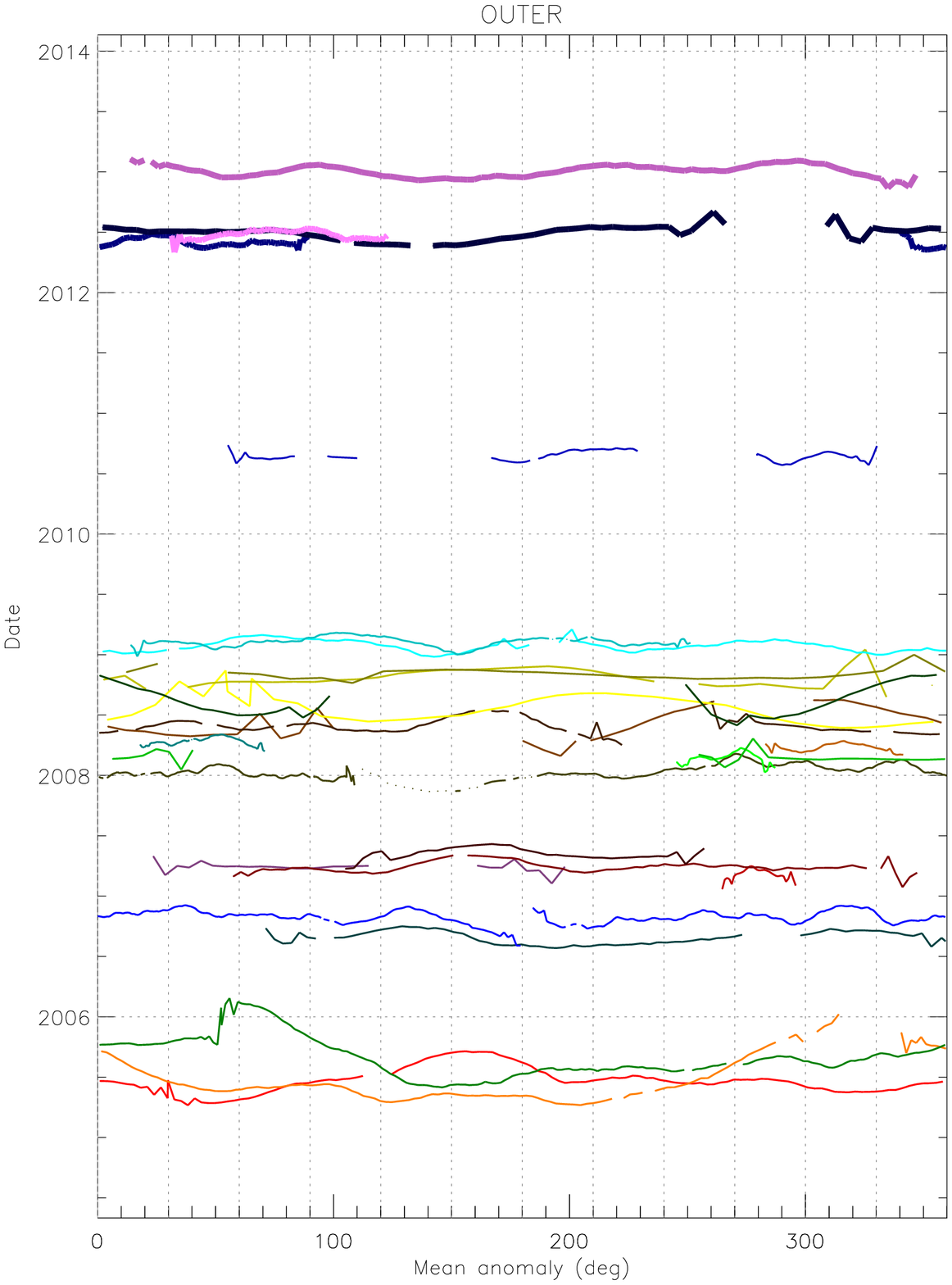}}}

{\scalebox{0.4}{\includegraphics{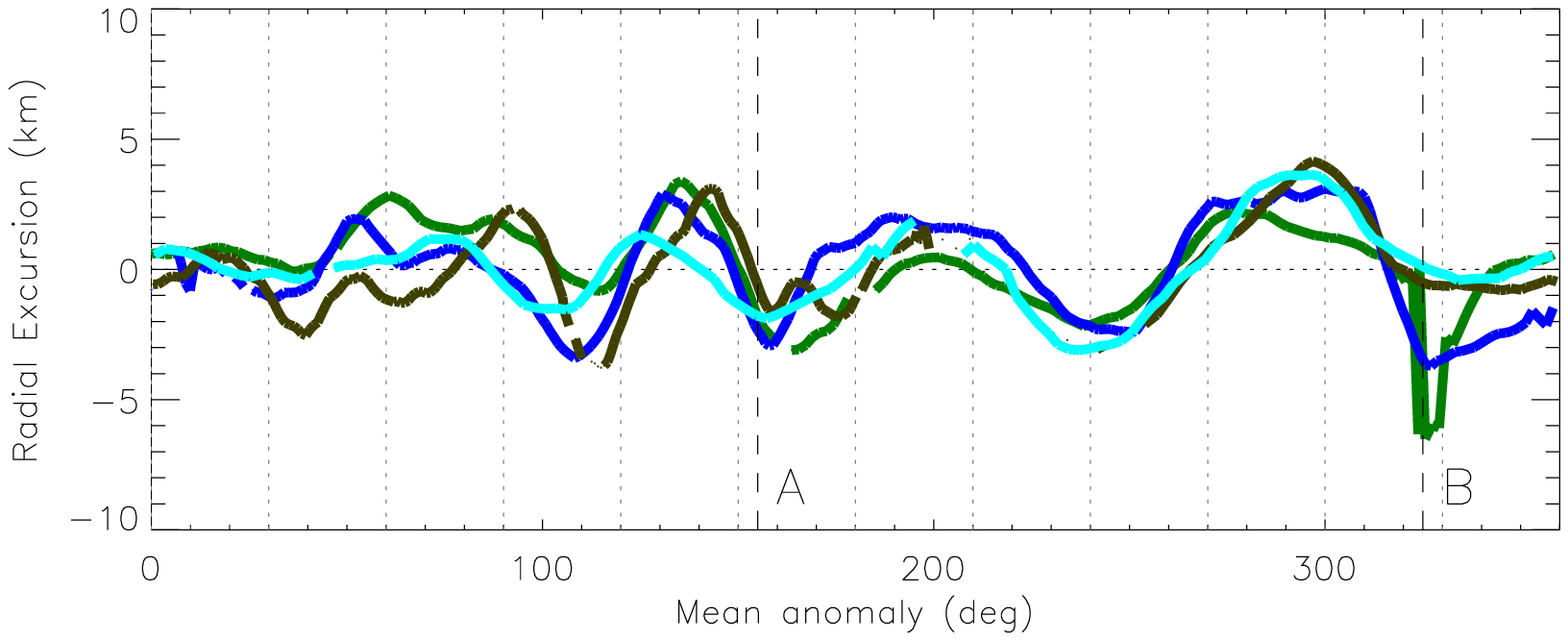}}}
{\scalebox{0.4}{\includegraphics{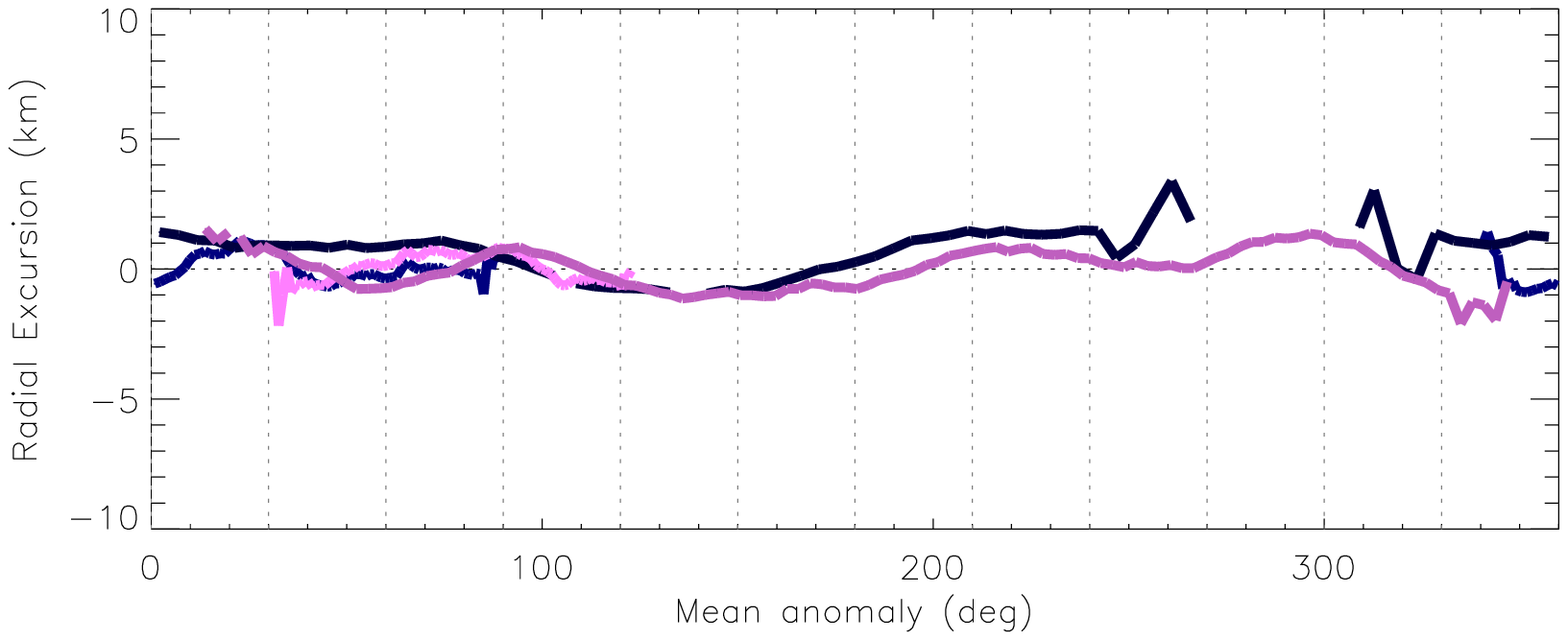}}}

 \caption{\small Residuals from the three-component fits (fit 744) plotted as radial excursion vs. mean anomaly.  Colors denote data sets as in in Fig. \ref{fig:m=1}.  Residuals were high-pass filtered by subtracting a 2nd-degree polynomial fit from each data set.  Each pattern was precessed at the speed expected for Keplerian orbits at that edge.  In the upper panels, the radial values are offset according to date.  In the lower panels, select curves (identified with thicker lines in the upper panels) are plotted together (with set 8 inverted on the left), demonstrating the coherence of an irregular pattern moving at that rate.  On the inner edge, the pattern is recognizable for at least 3.5 yr; on the outer edge, it is recognizable for several mo.  Labels A and B refer to features identified in Fig. \ref{fig:AB}.}
 \label{fig:keplerian}
\end{center}
\end{figure}
%--------------------------------------------------------------------------

The top panels in Fig. \ref{fig:keplerian} show residuals from the three-component fits for each data set, plotted by date, precessed at the speeds expected for Keplerian orbits at each edge.  The bottom panels show select residual profiles (indicated as thick curves in the upper panel) plotted over one another to demonstrate their congruity.  The bottom left panel shows that the distinct irregular shape on the inner edge persists and remains roughly fixed in a frame rotating at the expected Keplerian rate, though sometimes inverted, from the early data sets taken in the middle of 2005, through at least early 2009, a span of $\sim$3.5 yr.  Due to the irregular shape it can be seen that the patterns are unambiguously aligned during this interval.  Data sets obtained $\sim$4 yr later align well with one another over a period of nearly a year, but it is not clear whether features in those late data sets correlate with features in the earlier data sets because the pattern becomes more regular with time (i.e., the bumps in the late data sets all look about the same).  The relative inversion of the pattern at some times implies that the pattern oscillates radially and therefore reflects a perturbation in eccentricity rather than semimajor axis.  Scans that do not show a strong radial signature may have been taken near quadrature, though many scans suffer from poor filtering and provide ambiguous comparisons.  The recurrence of this coherent irregular pattern is not consistent with a superposition of unmodeled normal modes.  

Residuals on the outer edge are smaller than on the inner edge, and the filtering was even less effective in some scans, but analogous trends to those on the inner edge are evident among some scans, though it is difficult to recognize the pattern over timescales longer than several months, and the pattern is more regular. 

The difference in pattern speeds between the inner and outer edges of the ringlet is $\Delta n \sim (dn/da) \Delta a = \frac{3}{2}\sqrt{\mu/a^5}\Delta a$, so it takes an interval of $\Delta P = 2\pi/\Delta n$ to execute one synodic period.  For the Huygens ringlet, that period is a bit over 5 yr, so if the irregular component of the shape of the inner edge is driven by embedded masses, the objects responsible for the observed shape must occupy orbits in a narrow range of semimajor axis near the edge of the ring in order for the pattern to remain coherent for 3.5 yr.  Since little change is seen in the pattern during that interval, the synodic period of the bodies perturbing that edge must be well over an order of magnitude longer, or greater than 35 yr.  If so, then the perturbers must be distributed over a region within $\Delta a' = \frac{2}{3}\sqrt{a^5/\mu}\Delta n' \simeq$ 3 km or less of the ring edge.  

The smaller residuals on the outer edge suggest that any embedded perturbers are either smaller or further from the edge than is the case for the inner edge.  The shorter coherence timescale would argue in favor of perturbers spread over a wider zone interior to the ring edge, rather than smaller perturbers.  In other words, there are fewer large perturbers near the outer edge than near the inner edge.  On the other hand, the more regular pattern is less suggestive of embedded masses than the pattern on the inner edge.  

The persistent irregular pattern is not likely to arise from a superposition of unmodeled free normal modes moving at speeds given by Eq. \ref{eq:speeds}.  That equation implies that the pattern speed only approaches $n$ for very large values of the wave number, $m$.  Moreover, as a superposition of Fourier components cannot produce a signal with a higher frequency than any of the components, the observed pattern requires significant contributions from modes with speeds close to $n$, which would have very large wave numbers.  For example, a Fourier component whose speed differs from the Keplerian rate by one cycle in 35 yr would require a wave number of $m\sim$ 25000.  There is no known mechanism for producing a spectrum of unforced modes with preferentially large wave numbers in just the right combination of amplitudes and phases to produce a static pattern moving near the Keplerian rate.  On the other hand, perturbations from objects embedded near the edge of the ring would naturally have a spectrum with such properties.

%--------------------------------------------------------------------------
\begin{figure}[h!]
 \begin{center}
{\scalebox{0.25}{\includegraphics{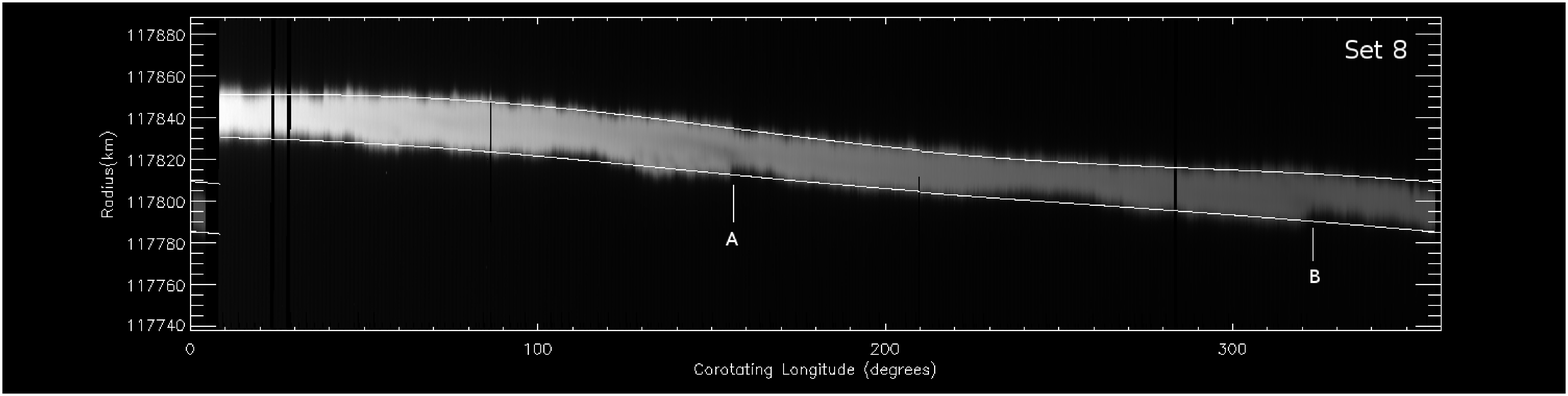}}}
{\scalebox{0.25}{\includegraphics{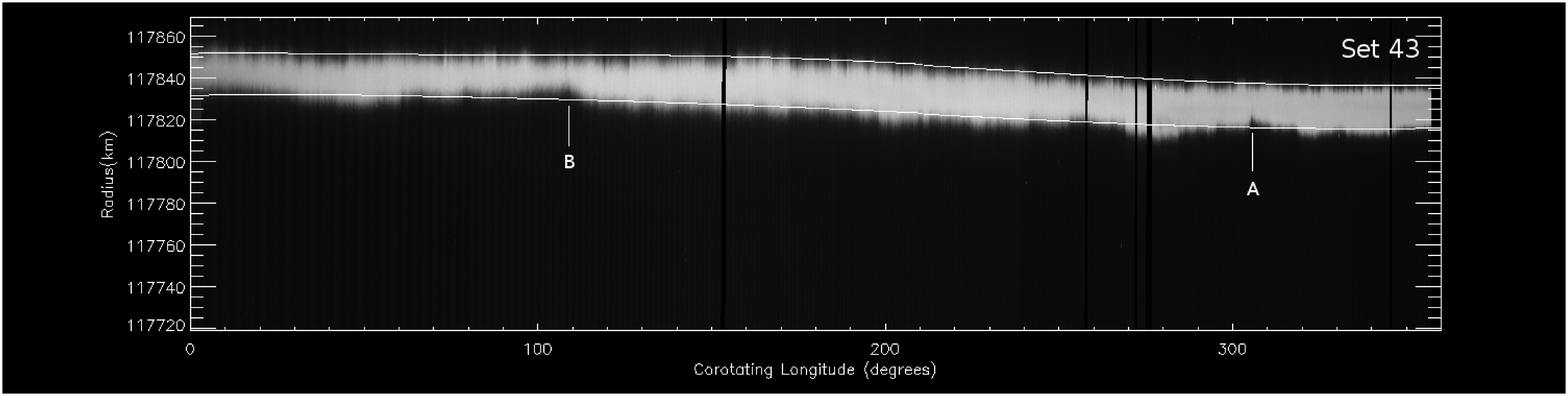}}}
 \caption{\small Mosaics in radius and azimuth for the two data sets with both high radial resolution and broad azimuthal coverage.  There are many irregularities in the ring edges, but features A and B, on the inner edge, indicate locations that appear to be the sources of satellite wakes.}
 \label{fig:AB}
\end{center}
\end{figure}
%--------------------------------------------------------------------------

Fig. \ref{fig:AB} shows further evidence of embedded masses at two locations near the inner edge of the ringlet.  These mosaics in radius and azimuth, for sets 8 and 43, were taken about 1 yr apart.  We chose those data sets because they had both high radial resolution and broad azimuthal coverage.  The features labeled A and B correspond to the mean anomalies with the same labels in Fig. \ref{fig:keplerian}.  The zeroes of the longitude system in Fig. \ref{fig:AB} were chosen to place the first and last images of each observation at the edges of the plots, and have no other significance.  In set 8, with the higher radial resolution, both features are associated with a dark wake-like structure consistent with the Keplerian shear direction.  The edge features are visible in set 43 as well, but the radial resolution is inadequate to resolve the wake structures.

%==================================================================
\section{Width-Radius Relation} 
\begin{figure}[h!]
 \begin{center}
{\scalebox{0.45}{\includegraphics{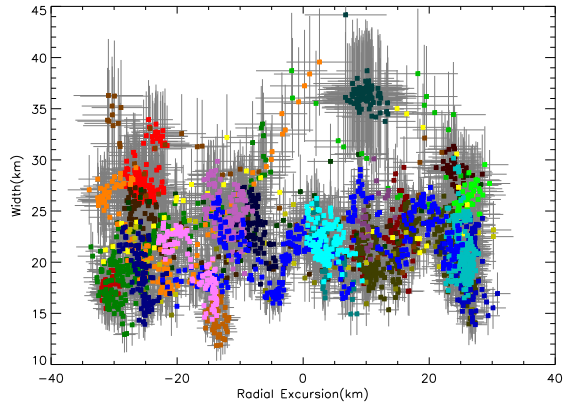}}}
 \caption{\small Width-radius relation for all data sets used in this study.  Colors denote data sets according to the legend in Fig. \ref{fig:m=1}.  No simple relationship is apparent.}
 \label{fig:wr-all}
\end{center}
\end{figure}

%==================================================================
\cite{Goldreich1979} showed that self gravity can maintain a narrow eccentric ringlet's apse alignment against differential apsidal precession if the eccentricity change $\delta e = e_2 - e_1$ across the ringlet is positive.  When the ringlet edges are perfectly aligned, and assuming $e << 1$, the width $W$ as a function of true anomaly $f$ is:
\begin{eqnarray}
	W(f) & = & a_2(1 - e_1 \cos f) - a_1(1 - e_1 \cos f) \nonumber \\  %
             & = & \delta a (1 - e \cos f) - a\delta e \cos f = %
	                          \delta a (1 - (e+q) \cos f), %
                                                                 \label{eq:b}
\end{eqnarray}
where $a=\frac{1}{2}(a_1+a_2)$, $e=\frac{1}{2}(e_1+e_2)$, and $q = a \delta e/\delta a$ is the eccentricity gradient.  Substituting $\cos f = (1-r/a)/e$ into (\ref{eq:b}), the width varies with radius as:
\begin{equation}
	W(r) = \delta a \left[1 - \left(e+\frac{q}{e}\right) %
                             \left(1-\frac{r}{a}\right) \right] . %
                                                                 \label{eq:c}
\end{equation}
The relation between the ringlet's width and mean radius should therefore be linear, with a slope $\gamma = e\delta a/a + \delta e/e$.  Since $e\delta a/a$ is always positive, the \cite{Goldreich1979} model requires $\gamma > 0$ for self gravity to maintain the alignment.  That condition is satisfied for the Uranian $\alpha$, $\beta$, and $\epsilon$ rings \cite{Elliott1983}, and for at least some of Saturn's narrow ringlets \cite{Spitale2006b}, but it is not satisfied for the Huygens ringlet.  Instead, as shown in Fig. \ref{fig:wr-all}, there does not appear to be a simple relationship between the ringlet's width and radius over the total interval that we examined, and the net slope of the widely scattered points is essentially zero.  That result is analogous to the \cite{Porco1983} result.

%--------------------------------------------------------------------------
%--------------------------------------------------------------------------

%--------------------------------------------------------------------------
\begin{figure}[h!]
 \begin{center}
{\scalebox{0.3}{\includegraphics{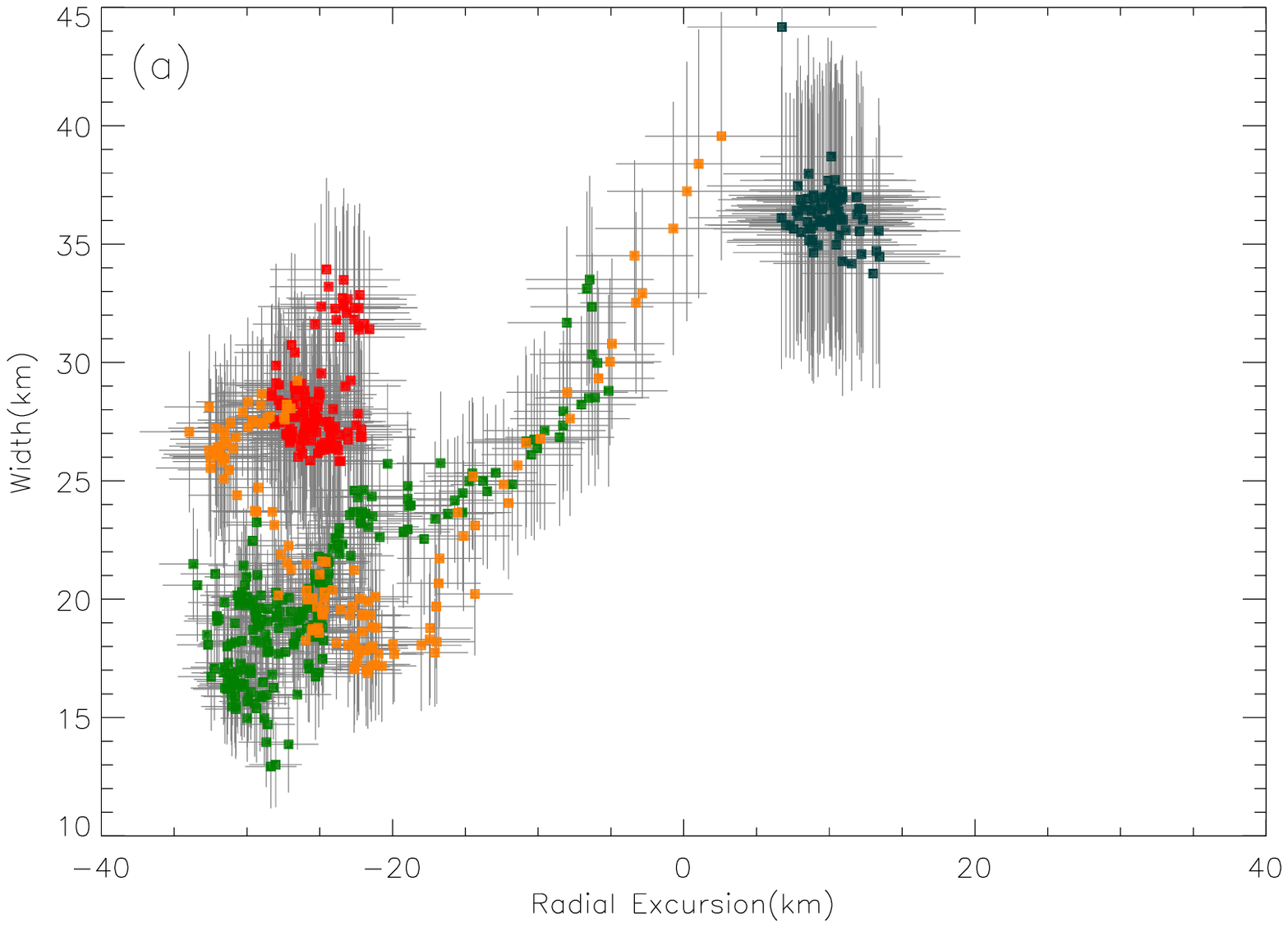}}}
{\scalebox{0.3}{\includegraphics{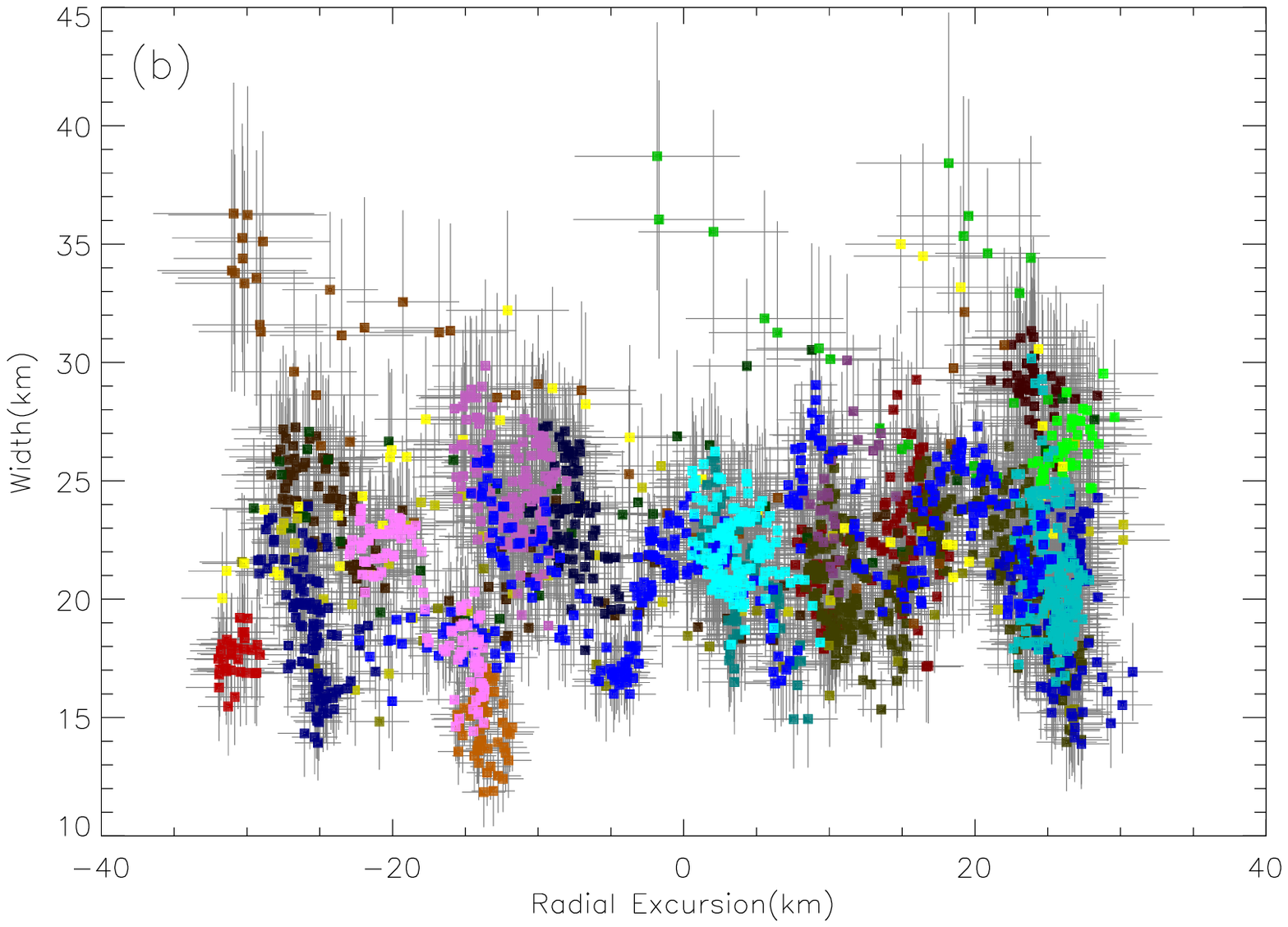}}}

{\scalebox{0.3}{\includegraphics{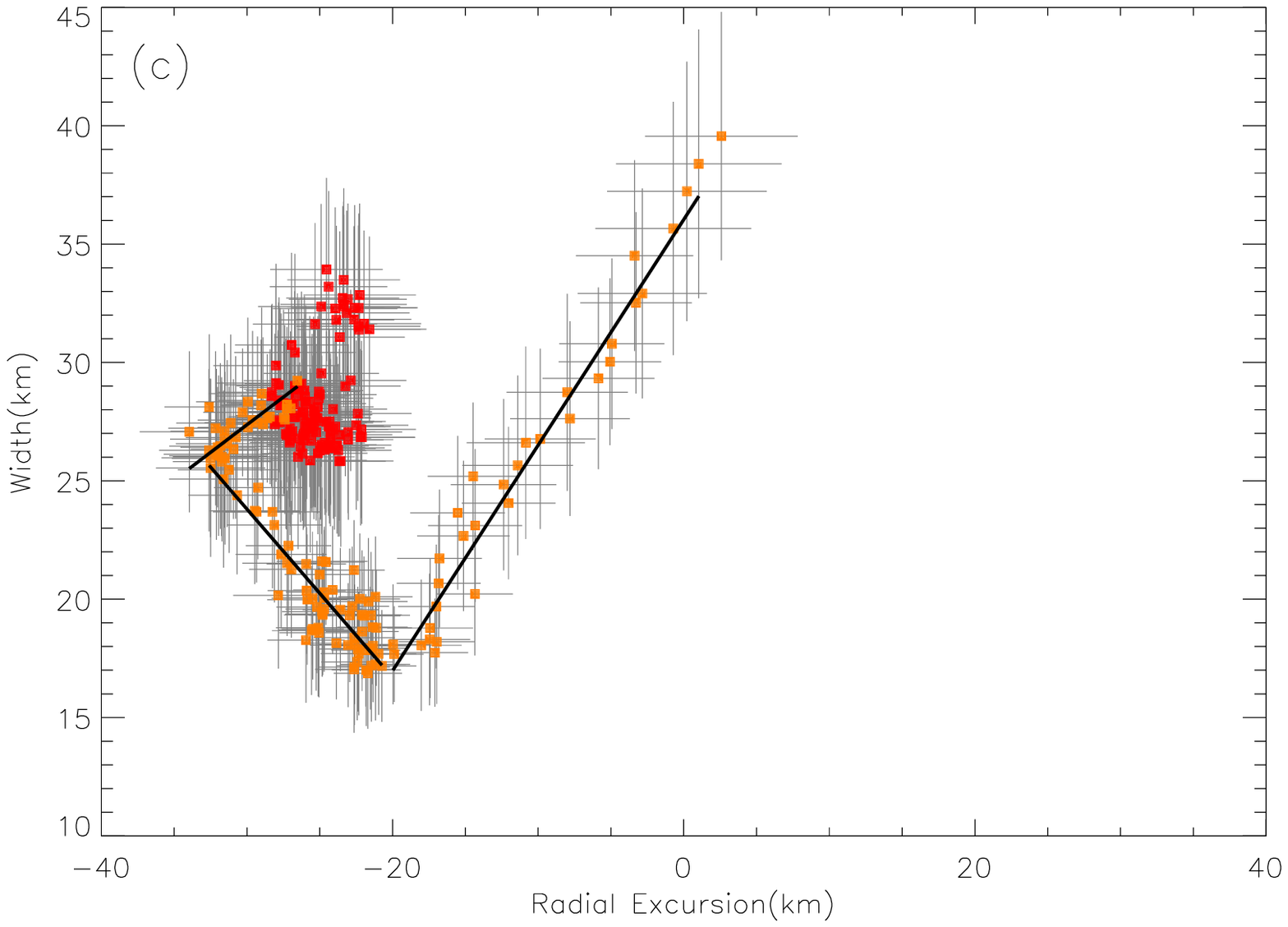}}}
{\scalebox{0.3}{\includegraphics{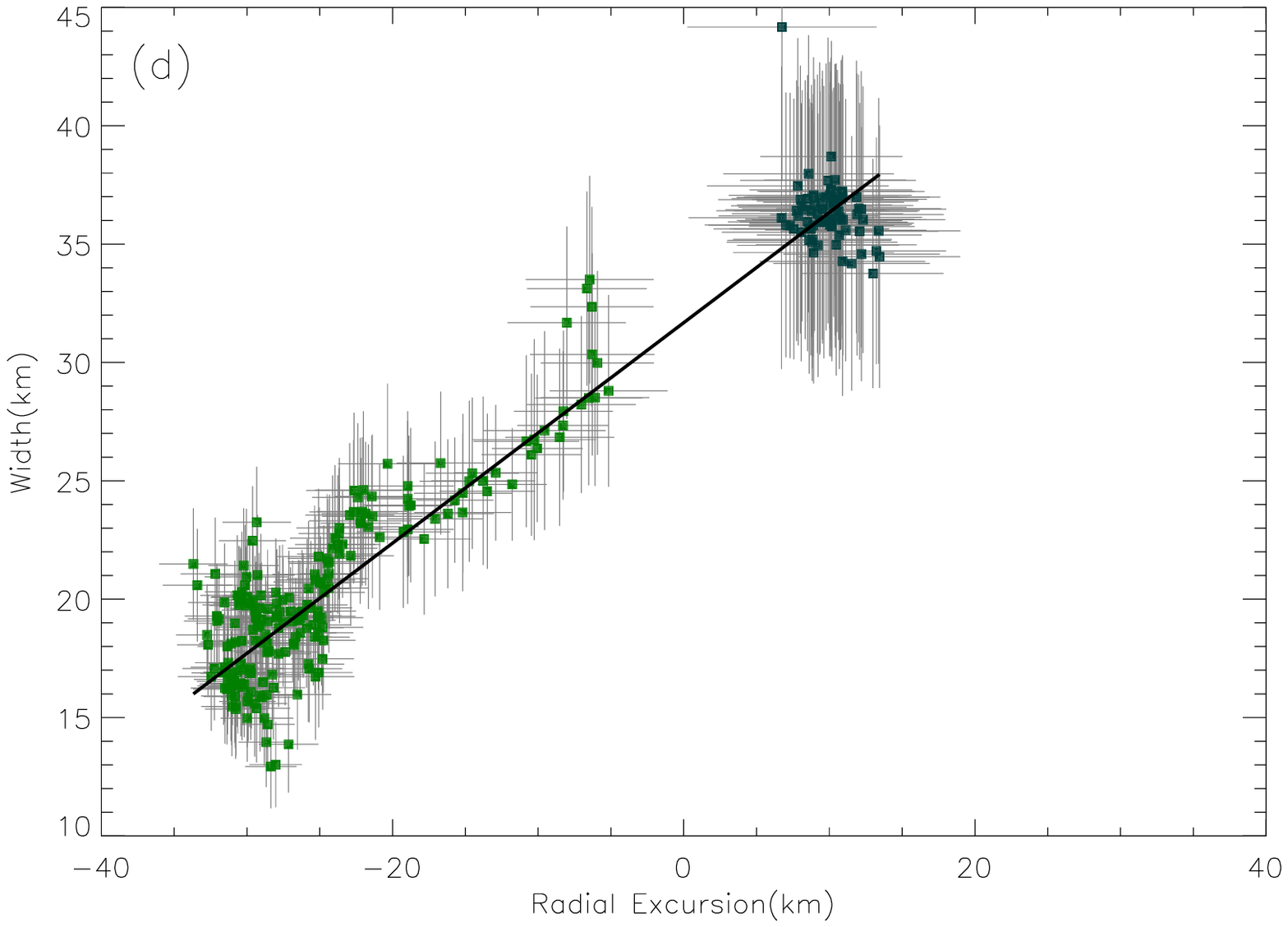}}}

 \caption{\small (a) Width-radius relation for sets 2, 3, 5, and 6.  (b) Width-radius relation for sets all other data sets.  (c) Piece-wise linear fit to apparent trends in the width-radius relation for sets 2 and 3.  (d) Linear fit to the width-radius relation for sets 5 and 6.  Colors denote data sets according to the legend in Fig. \ref{fig:m=1}.}
 \label{fig:wr-groups}
\end{center}
\end{figure}
%--------------------------------------------------------------------------

However, relatively simple behavior is seen during a shorter interval in the early data sets.  In the top row of Fig. \ref{fig:wr-groups}, the data sets are divided into "early" (a) and "late" (b) groupings, with the division occurring between 2006-247 (set 6) and 2006-312 (set 8).  There is a clear difference in the character of the width-radius relation between the early and late groups: the early group shows a steep overall slope with relatively low scatter, while the later group shows a much flatter slope and significant scatter.

The early group is further divided in the bottom row of Fig. \ref{fig:wr-groups}.  Sets 2 and 3 were taken about two days apart, and sets 5 and 6 were taken about two weeks apart, nearly two months later than sets 2 and 3.  Set 3 shows two linear trends with slopes of about 1 and -0.7, yielding a local minimum in width.  Set 3 also shows a hint of an additional linear trend with slope of about 0.5, which is consistent with set 2.  Sets 5 and 6 show a simple linear trend with a slope of about 0.5.

%--------------------------------------------------------------------------
\begin{figure}[h!]
 \begin{center}
{\scalebox{0.25}{\includegraphics{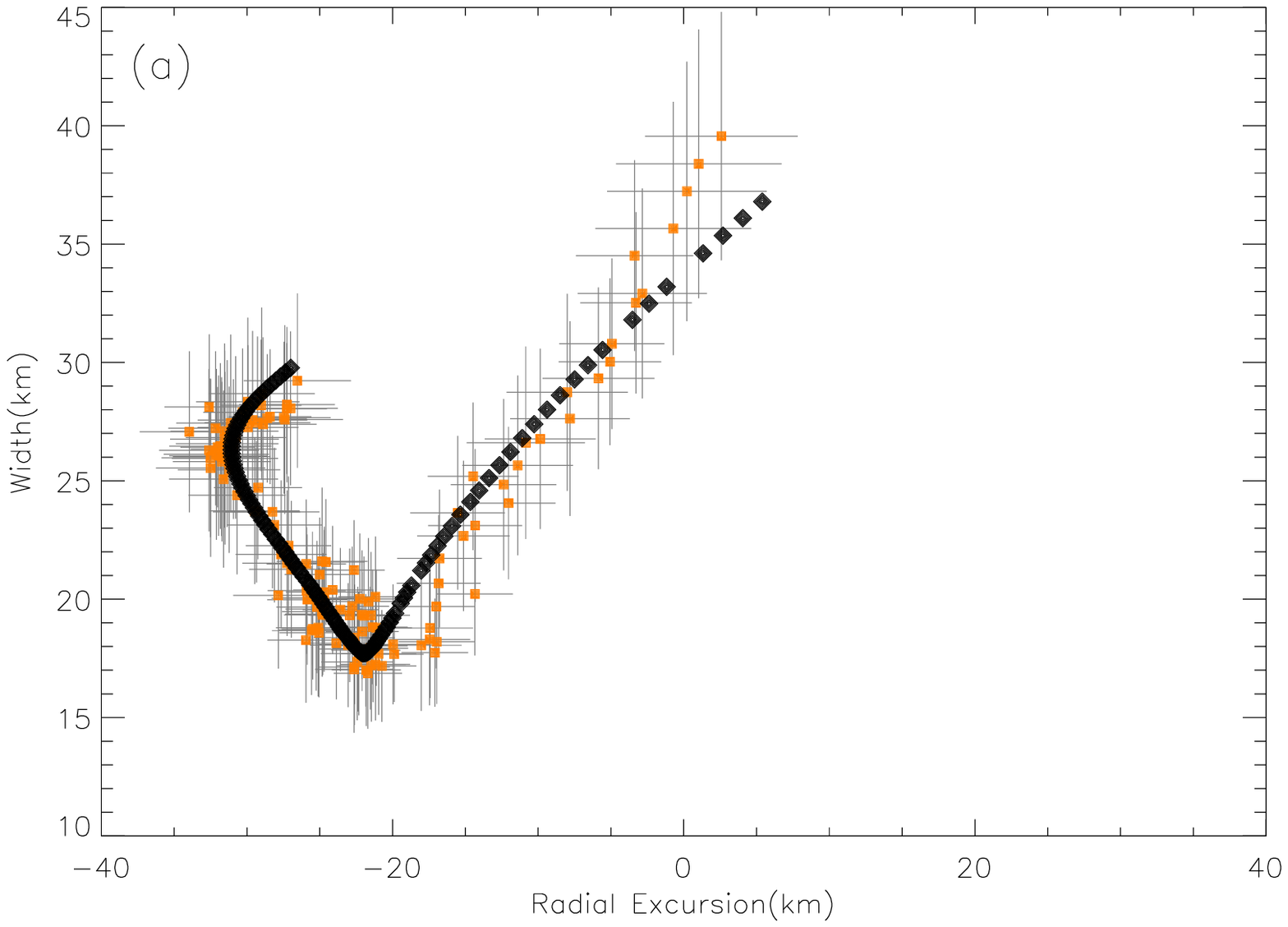}}}
{\scalebox{0.25}{\includegraphics{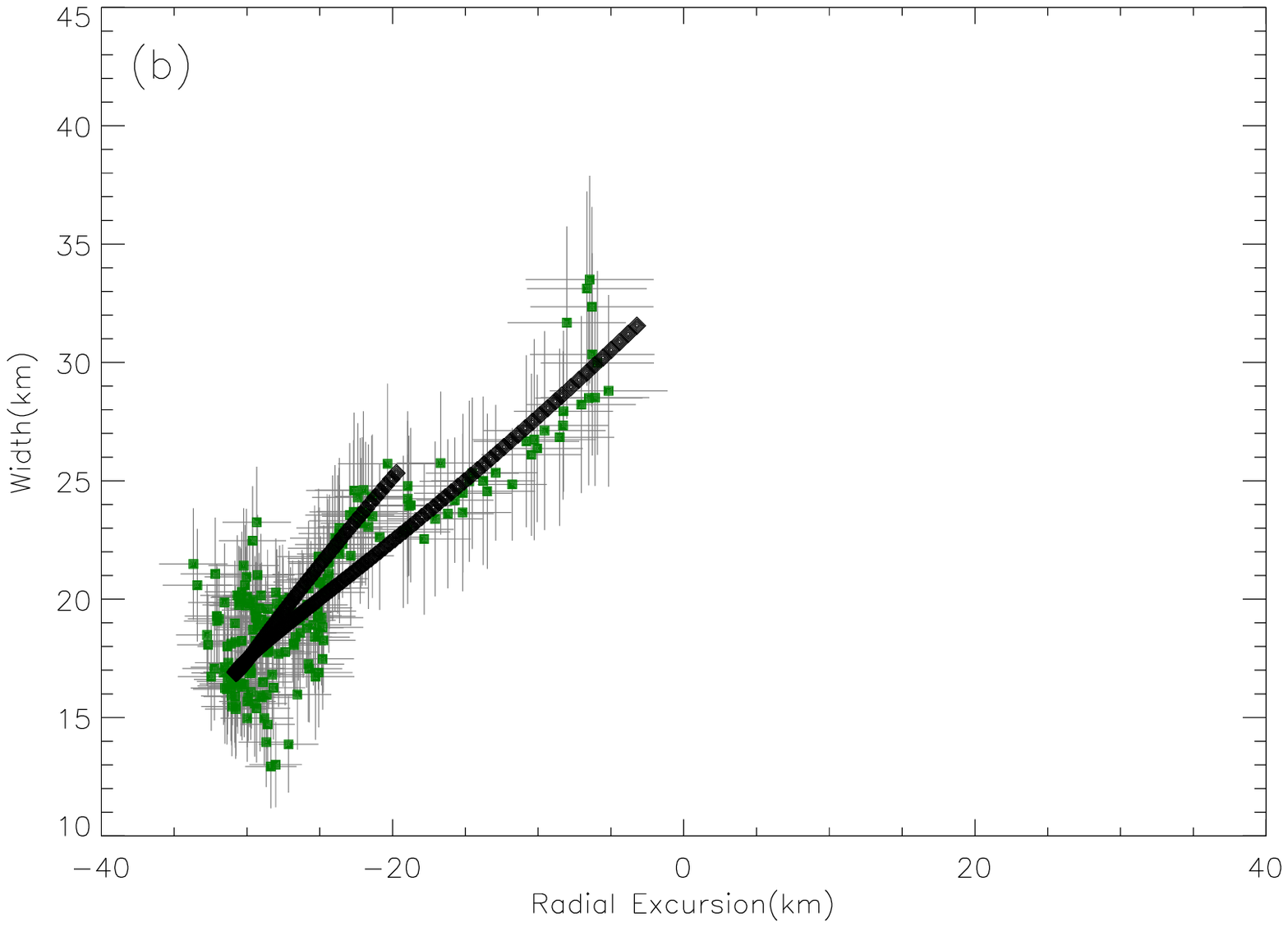}}}
{\scalebox{0.25}{\includegraphics{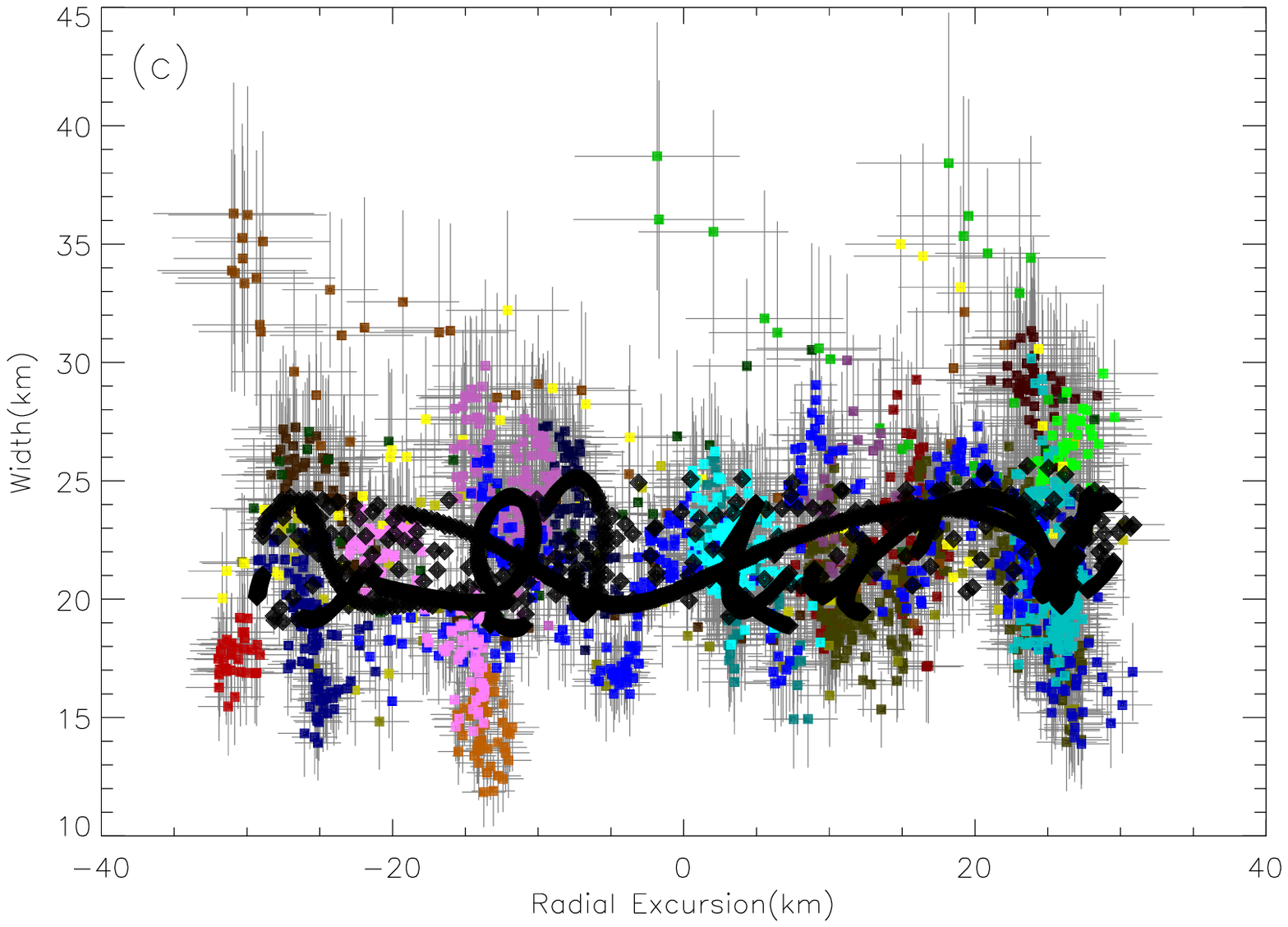}}}
 \caption{\small Comparing measured width-radius relations with those computed from fit 744.  (a) Sets 2 and 3.  (b) Sets 5 and 6.  (c)  All other data sets.  Colors denote data sets according to the legend in Fig. \ref{fig:m=1}.}
 \label{fig:wr-model}
\end{center}
\end{figure}
%--------------------------------------------------------------------------

The presence of a local minimum in width in Fig. \ref{fig:wr-groups}(c) suggests a significant apsidal misalignment at that time, even though it was not possible to directly measure the periapse longitudes for those data sets (see Sec. \ref{sec:variability}).  Fig. \ref{fig:wr-model} compares widths vs. radii for those fits (for clarity, the ansa movies -- sets 2 and 6 -- are excluded because they don't contribute much, as they occupy small regions in width-radius space), as well as for a three-component ($m=1,2^\mathrm{(forced)}, 2^\mathrm{(free)}$) fit to just the late data sets, with the observed widths and radii.  For a simple eccentric ringlet, an apsidal misalignment would produce a smooth ellipse-like curve in width-radius space.  The structure in the plotted model curves arises from the normal modes on the ring edges.  Significant deviations from the modeled curves are due to the irregular structure (see Sec. \ref{sec:irregular}) that is not modeled in the normal-mode fits.  The large scatter in the late plot is consistent with the trend of generally larger residuals seen on the inner edge in the later data sets in Fig. \ref{fig:keplerian}.

The time-averaged eccentricity gradient, $q = a \delta e/\delta a$, of about 0.008 derived from fit 744 is much smaller than for the Colombo and Maxwell ringlets determined by \cite{Porco1983}, which were both about 0.46.  Although the presence of $m \ne 1$ modes complicates the width-radius relation \cite{Longaretti1989}, the $m=1$ amplitude is still an order of magnitude greater than the amplitudes of the $m \ne 1$ modes.  Therefore, the essentially flat slope, and correspondingly tiny eccentricity gradient, implies that the shape of the Huygens ringlet is not continuously maintained by simple self gravity, like other narrow eccentric ringlets.

%==================================================================
\section{Discussion}  \label{sec:discussion}
%==================================================================
Our modeling corroborates, and builds upon, the broad results of \cite{Porco1983}: the primary time-averaged shape of the Huygens ringlet is a Keplerian ellipse with an amplitude of about 28 km, with inner and outer apses nearly aligned.  The width-radius relation is complicated, demonstrating that the kinematics of the Huygens ringlet are fundamentally different than for most other known narrow eccentric rings.

Mode searches unambiguously revealed a forced (by the Mimas resonance) $m=2$ mode on the inner edge and a free $m=2$ mode on the outer edge of the ringlet.  No evidence for modes corresponding to forcing from the B-ring edge was seen.  A local maximum in "significance" (see Sec. \ref{sec:modes}) was seen for a forced mode on the outer edge as well.  Examination of the variation of the total $m=2$ amplitude with time (Fig. \ref{fig:amplitudes}) supports the presence of the forced $m=2$ mode on the outer edge, and a direct fit to a model containing that mode shows a statistically significant pattern.  Therefore, we conclude that the forced mode on the outer edge is real, but with a smaller amplitude than the single-particle theory would predict.  The large phase lag observed for that mode is consistent with significant dissipation, which would reduce the amplitude from the predicted value.

Each edge of the Huygens ringlet possesses an irregular shape component that is not explained by our normal mode modeling.  The irregular component moves at a speed near the local Keplerian rate and is coherent for several months on the outer edge and several years on the inner edge.  The pattern is easier to characterize on the inner edge because of the larger amplitudes of the residuals from the three-component normal-mode fit.  The pattern reflects a perturbation primarily in eccentricity rather than semimajor axis (see Sec. \ref{sec:irregular}).  It is implausible that a superposition of unmodeled free normal modes would happen to give rise to such a static pattern, so we consider the most likely cause to be perturbations from bodies embedded in the ring.  Assuming each object produces a localized perturbation that moves at the same orbital rate as the object, the objects that produce the irregular shape must reside within a few km of the edge; otherwise, the observed pattern would evolve much faster than observed.  The large vertical thickness of the Huygens ringlet compared to the Maxwell and Colombo ringlets (See Sec. \ref{sec:m=2}) may be a reflection of stirring from such embedded fragments.

Long wake-like structures originate from two locations on the inner edge of the ringlet, likely revealing the locations of the two largest embedded masses.  Those features look like the top half of the kind of propeller-shaped disturbances typically associated with embedded perturbers.  However, they span about one half of the 20-km width of the ring, in contrast with the largest propellers in the A-ring, whose radial extents are not much larger than 5 km \cite{Tiscareno2010}.  For such propellers, the radius of the perturber has been estimated as $R \sim 0.18 \Delta r$, where $\Delta r$ is the radial extent of each lobe of the propeller \cite{Seiss2005, LewisStewart2009,Porco2007}.  The radial extent of the larger feature, A, is about 10 km, yielding a diameter estimate of $R \sim$ 3.6 km.  Such a body might be resolvable in some imaging sequences against a dark sky, but special lighting may be required to see a body with the ring as a backdrop.  Therefore, the fact that we have not identified any such bodies in the images does not rule out their existence.

It is worth commenting on the resemblance of the Huygens ringlet to the F ring.  Both ringlets show the effects of embedded bodies \cite{Spitale2006, Murray2005, Murray2008}, though the morphology and dynamics of the F ring are dominated by the presence of massive bodies to a greater degree than apparently is the case for the Huygens ringlet.  That is not surprising, given the F ring's location near Saturn's Roche limit, but that explanation does not apply to the Huygens ringlet.  Moreover, the very existence of the Huygens ringlet is unexplained, as there are no apparent shepherds to confine it, as in the case of the F ring.  If the Huygens ringlet is the remnant of the disruption of a small satellite, then the embedded bodies affecting the shape of the ringlet would represent the largest fragments from that event.  In that case, the ringlet is a transient feature whose lifetime would depend on its viscous evolution.

Except for the observations taken prior to late 2006, the Huygens ringlet generally shows a flat width-radius relation, and a correspondingly small eccentricity gradient, though for those few early observations, there is a significant eccentricity gradient and the ringlet does look like a self-gravitating narrow ringlet \cite{Goldreich1979}.  Therefore, although the ringlet may occasionally be configured to maintain its alignment via self gravity, that configuration does not appear to be typical.  Given the $\sim$ 300-yr synodic period between the precession of the inner vs. outer edges of the ringlet, the ringlet would need to correct its apsidal alignment at least every few decades to avoid significant degradation of the $m=1$ pattern, assuming the ringlet is even that old.  However, Fig. \ref{fig:dlp} shows that the relative alignment was not simply drifting, but may have been librating, during the latter half of 2008, a time during which the eccentricity gradient was not configured to maintain the alignment via self gravity.  Libration would imply that some other mechanism is responsible for maintaining the alignment, or that more than just self gravity is required \cite{ChiangGoldreich2000}.    

The dramatic change in character of the width-radius relation in 2006 raises the question of what may have happened to the Huygens ringlet around that time.  The ringlet was changing rapidly prior to that time, so any event may have occurred before our observations began.  Voyager observed a similarly disturbed ringlet $\sim$1/4 century earlier, so at least two putative episodes have been observed.  Moreover, viscosity would be expected to have damped the observed apsidal librations \cite{Borderies1982}.  If the disturbances in the Huygens ringlet were produced by external agents, the most likely candidate would be impacts.  In order to scramble the kinematics of the entire ringlet an impactor would likely need a broad physical extent, and would therefore likely consist of an extended cloud of material, rather than a solid body, similar to the conditions required to produce the vertical corrugations seen elsewhere in Saturn's rings \cite{Hedman2011,Hedman2015}, and in Jupiter's rings \cite{Showalter2011}.  However no other evidence has come to light regarding disturbances in nearby ring material, requiring the coincidence of two direct hits on the Huygens ringlet by extended, but not too extended, objects during the Voyager and Cassini observations.  

A simpler explanation would be that the disturbances in the Huygens ringlet reflect a process inherent to the ringlet.  That process must be periodic (or at least episodic) and the period must be longer that the $\sim$8-yr time span of the observations.  The beat period of the total $m=2$ pattern in the B-ring is $\sim$5 yr \cite{Spitale2010}, so a direct response to the strongest forcing from the B-ring edge is ruled out (consistent with the results from the normal mode modeling).  If the period is instead tied to the synodic (or something like a horseshoe) period of the largest embedded masses, a longer baseline of observations will be required for confirmation.  In any case, the presence of the embedded masses likely complicates the ringlet's kinematics.

%==========================================================
\section{Acknowledgements} 
%==========================================================
We acknowledge P. D. Nicholson, R. G. French, and P. Goldreich for helpful discussions regarding the presence of higher-order normal modes.  We thank R. A. Jacobson for sharing his fits to Mimas' mean motion.  This paper was greatly improved by the thorough reviews submitted by two anonymous referees.  This work was supported under grant number NNX11AP73G of the NASA Outer Planets Research program.  JMH's efforts here were supported by the National Science Foundation via Grant No. AST-1313013, and JMH also thanks Byron Tapley for graciously providing office space and the use of the facilities at the University of Texas Center for Space Research (CSR).

\pagebreak
%=====================================================================

%=====================================================================

\end{document}